\documentclass{aa}  
\usepackage{graphicx}
\usepackage{natbib}
\bibpunct{(}{)}{;}{a}{}{,} 
\usepackage{txfonts}

\begin{document}

   \title{The stellar-to-halo mass relation over the past 12 Gyr}

   \subtitle{I. Standard $\Lambda$CDM model}

   \author{G. Girelli
          \inst{1,2}
          \thanks{giacomo.girelli2@unibo.it}
          \and
          L. Pozzetti\inst{1}
          \and
          M. Bolzonella\inst{1}
          \and
          C. Giocoli\inst{2,1,3,4}
          \and
          F. Marulli\inst{2,1,3}
          \and
          M. Baldi\inst{2,1,3}
          }

   \institute{INAF – Osservatorio di Astrofisica e Scienza dello Spazio di Bologna, via Gobetti 93/3, I-40129 Bologna, Italy
         \and
             Dipartimento di Fisica e Astronomia, Universit\`a di Bologna, via Gobetti 93/2, I-40129, Bologna, Italy
             \and
             INFN – Sezione di Bologna, viale Berti Pichat 6/2, I-40127 Bologna, Italy
             \and
             Dipartimento di Fisica e Scienze della Terra, Universit\`a degli Studi di Ferrara, via Saragat 1, I-44122 Ferrara, Italy
             }

   \date{Received XXX; accepted YYY}

 
  \abstract
    {}
   {Understanding the link between the galaxy properties and the dark matter halos they reside in and their coevolution is a powerful tool for constraining the processes related to galaxy formation. In particular, the stellar-to-halo mass relation (SHMR) and its evolution throughout the history of the Universe provides insights on galaxy formation models and allows us to assign galaxy masses to halos in N-body dark matter simulations. To address these questions, we determine the SHMR throughout the entire cosmic history from $z \sim 4$ to the present.}
   {We used a statistical approach to link the observed galaxy stellar mass functions on the COSMOS field to dark matter halo mass functions up to $z \sim 4$ from the $\Lambda $CDM {\small DUSTGRAIN-}{\it pathfinder} simulation, which is complete for $M_h>10^{12.5}M_\odot$ , and extended this to lower masses with a theoretical parameterization. We propose an empirical model to describe the evolution of the SHMR as a function of redshift (either in the presence or absence of a scatter in stellar mass at fixed halo mass), and compare the results with several literature works and semianalytic models of galaxy formation. We also tested the reliability of our results by comparing them to observed galaxy stellar mass functions and to clustering measurements.}
   {We derive the SHMR from $z=0$ to $z=4$, and model its empirical evolution with redshift.  We find that $M_*/M_h$ is always lower than $\sim 0.05$ and depends both on redshift and halo mass, with a bell shape that peaks at $M_h\sim10^{12}\,M_\odot$. Assuming a constant cosmic baryon fraction, we calculate the star-formation efficiency of galaxies and find that it is generally low; its peak increases with cosmic time from $\sim 30\%$ at $z\sim 4$ to $\sim 35\%$ at $z\sim 0$. Moreover, the star formation efficiency increases for increasing redshifts at masses higher than the peak of the SHMR, while the trend is reversed for masses lower than the peak. This indicates that massive galaxies (i.e., galaxies hosted at halo masses higher than the SHMR peak) formed with a higher efficiency at higher redshifts (i.e., downsizing effect) and vice versa for low-mass halos. We find a large scatter in results from semianalytic models, with a difference of up to a factor $\sim 8$ compared to our results, and an opposite evolutionary trend at high halo masses. By comparing our results with those in the literature, we find that while at $z\sim 0$ all results agree well (within a factor of $\sim 3$), at $z>0$ many differences emerge. This suggests that observational and theoretical work still needs to be done. Our results agree well (within $\sim 10\%$) with observed stellar mass functions (out to $z=4$) and observed clustering of massive galaxies  ($M_*>10^{11}M_\odot$ from $z\sim 0.5$ to $z\sim 1.1$) in the two-halo regime.}
{}

   \maketitle
%

\section{Introduction}
In the current $\Lambda$ cold dark matter ($\Lambda$CDM) concordance cosmology, the matter density of the Universe is dominated by CDM, whose gravitational evolution gives rise to a population of virialized dark matter halos spanning a wide range of masses. Numerical simulations of structure formation in a CDM universe predict that these dark matter halos contain a population of subhalos. In this picture, galaxies form at the centers of halos and subhalos, and their formation is mainly driven by the cooling and condensation of gas in the center of the potential wells of the extended virialized dark matter halos \citep{White78}. Therefore, galaxy properties such as luminosity or stellar mass are expected to be tightly coupled to the depth of the halo potential and thus to the halo mass. Understanding the relation of the galaxy stellar mass content to the mass of its dark matter halo is a powerful tool for constraining the processes related to galaxy formation.

There are many different approaches to link the properties of galaxies to those of their halos (see \citealt{Wechsler18} for a review). A first method derives the halo properties from the properties of its galaxy population using galaxy kinematics \citep{Erickson87,More09,More09B,More11,Li12}, gravitational lensing \citep{Mandelbaum05,Velander14}, or X-ray studies \citep{Lin03,Lin04,Kravtsov18}, for instance.
A second approach is to model the physics that shapes galaxy formation in either large numerical simulations that include both gas and dark matter (\citealt{Springel03}, or the Illustris simulation \citealt{Sijacki15,Nelson15} or the Eagle simulation \citealt{McAlpine16}) or semianalytic models (SAMs) of galaxy formation (e.g., \citealt{Kauffmann93, Gonzalez14, Henriques15, Croton16}). However, many of the physical processes involved in galaxy formation (such as star formation and feedback) are still not well understood, and in many cases, simulations are not able to reproduce the observed quantities with high accuracy (see \citealt{Naab17,Somerville15} for two reviews) over large volumes. Building realistic and large mocks is fundamental for optimizing the scientific exploitation of ongoing and future large surveys, such as Euclid \citep{Laureijs11}, by means of understanding and minimizing the systematic uncertainties and selection effects.

With the advent of large galaxy surveys in the past decades, other methods have been developed. These link galaxies to halos using a statistical approach. One example is the halo occupation distribution (HOD) formalism, which specifies the probability distribution for a halo of mass $M$ to harbor $N$ galaxies with certain intrinsic properties, such as luminosity, color, or type (e.g., \citealt{Zheng07, Yang12, Carretero15}). More complex formulations of this type of modeling, such as the conditional luminosity function (CLF) formalism (e.g., \citealt{Vandenbosch03, Yan05, Yang12}), have extended the HOD approach. Because reliable galaxy clustering measurements are not available at high redshift, the HOD and CLF approaches have typically been used only at low redshift.
In order to avoid this problem, galaxies and dark matter halos can be linked assuming that a galaxy property (i.e.,~the stellar mass, or the galaxy luminosity) monotonically relates to a halo property (i.e., the halo mass, the circular velocity of halos), and therefore the relation between dark matter halos or subhalos and galaxy properties can be found by performing a one-to-one association from their distributions. This approach is called (sub)halo abundance matching, hereafter SHAM, or simply AM (e.g.,~\citealt{Guo10, Behroozi10, Behroozi13, Reddick13, Moster13, Rodriguez17}). The only observational input of this method is the stellar mass function (SMF) or luminosity function. This approach also predicts the clustering statistics remarkably well (see \citealt{Moster10,Moster13} as examples), down to scales that depend on the resolution of the adopted simulation (an additional modeling of subhalos is required on scales smaller than the resolution).
The evolution of the SHAM technique is the empirical modeling (EM, e.g., \citealt{Behroozi19,Moster18,Grylls19,Schreiber17}), in which dark matter halos from an N-body simulation (with merger trees) are linked to several observed galaxy properties (such as stellar mass, star-formation rate, and quenched galaxy fractions).

These methods have the advantage that they do not rely on assumptions about the many details of physical processes that drive galaxy formation. Moreover, by construction, physical properties are in agreement with observations, and the parameterization of the model is flexible (given that the parametrization only depend on the input observations and can easily be changed) but this type of modeling is less predictive than other types, such as hydrodynamic simulations or SAMs \citep{Wechsler18}.
However, these models can constrain the relationship between galaxy and halo properties (and thus, indirectly, the underlying physics), and mock catalogs can be constructed that reproduce  an observational quantity in detail (such as the SMF).

We here adopt the SHAM approach to link the SMF to halo and subhalo mass functions. In particular, we use a dedicated SMF derived on the Cosmological Evolution Survey (COSMOS) field \citep{Scoville07}, whose size and depth allow us to apply this method over a wide homogeneous and continuous redshift coverage up to $z\sim4$. We describe the cosmological simulation we use in Sect.~\ref{simulations} and the observations in Sect.~\ref{observedSMF}. In Sect.~\ref{SHMR} we present our method,  parameterize the stellar-to-halo mass relation (SHMR), and model its empirical evolution with redshift. The results are presented in Sect.~\ref{sectSHMR}, with an analysis of the evolution of the SHMR peak and its implications for the physics of galaxy formation and evolution.
In order to test the validity of our method, in Sect.~\ref{testingMF} we apply the relation to the dark matter simulation; we then compute the SMFs, and compare them with different observed estimates. Finally, in Sect.~\ref{clustering} we analyze clustering as a function of stellar mass. 

Throughout this paper we adopt a standard $\Lambda$CDM cosmology with cosmological parameters set to be consistent with the Planck 2015 constraints \citep{Planck16}: $\Omega_m = 0.31345$, $\Omega_\Lambda = 0.68655$, $\Omega_b=0.0481$,  $H_0= 67.31\,\mathrm{km}\, \mathrm{s}^{-1}\,\mathrm{Mpc}^{-1}$, and $n = 0.9658$. More importantly, we highlight that all the masses reported in this paper (both halo and stellar masses, unless differently specified) are expressed in units of $h_{67}= H_0\,/(67\,\mathrm{km}\, \mathrm{s}^{-1}\,\mathrm{Mpc}^{-1})$. More precisely, halo masses are in units of $M_h\,h^{-1}_{67}$, and stellar masses are in units of $M_*\,h^{-2}_{67}$.


\section{Simulation and halo catalogs}\label{simulations}

In this section we present the simulation we adopt in this work. We describe its parameters, the halo catalogs, and the adopted halo masses.

\subsection{N-body simulation}

The populations of dark matter halos we used were drawn from a cosmological N-body collisionless simulation run with the code \textsc{MG-GADGET} \citep{Springel01, Springel05, Puchwein13} within the {\small DUSTGRAIN-}{\it pathfinder} simulation set presented in \citet{Giocoli18}. Standard cosmological parameters were set to be consistent with the Planck 2015 cosmic microwave background (CMB) based cosmological constraints \citep{Planck16} mentioned above.
The simulation was performed in a periodic cosmological box of $750\,\mathrm{Mpc}\,h^{-1}$ per side, and contained $768^3$ particles with a particle mass of $m_{\rm CDM} = 8.1\,\times\, 10^{10}\,M_\odot\,h^{-1}$, with $h = H_0\,/(100\, \mathrm{km}\,\mathrm{s}^{-1}\,\mathrm{Mpc}^{-1})$.
We chose the $\Lambda $CDM {\small DUSTGRAIN-}{\it pathfinder} as a reference simulation although its mass resolution is quite low because we will extend the approach described in this paper to modified gravity and/or massive neutrino cosmologies, which are present in the {\small DUSTGRAIN-}{\it pathfinder} suite.

\subsection{Halo catalogs}\label{halocat}

\begin{figure}
   \centering
   \includegraphics[width=0.49\textwidth]{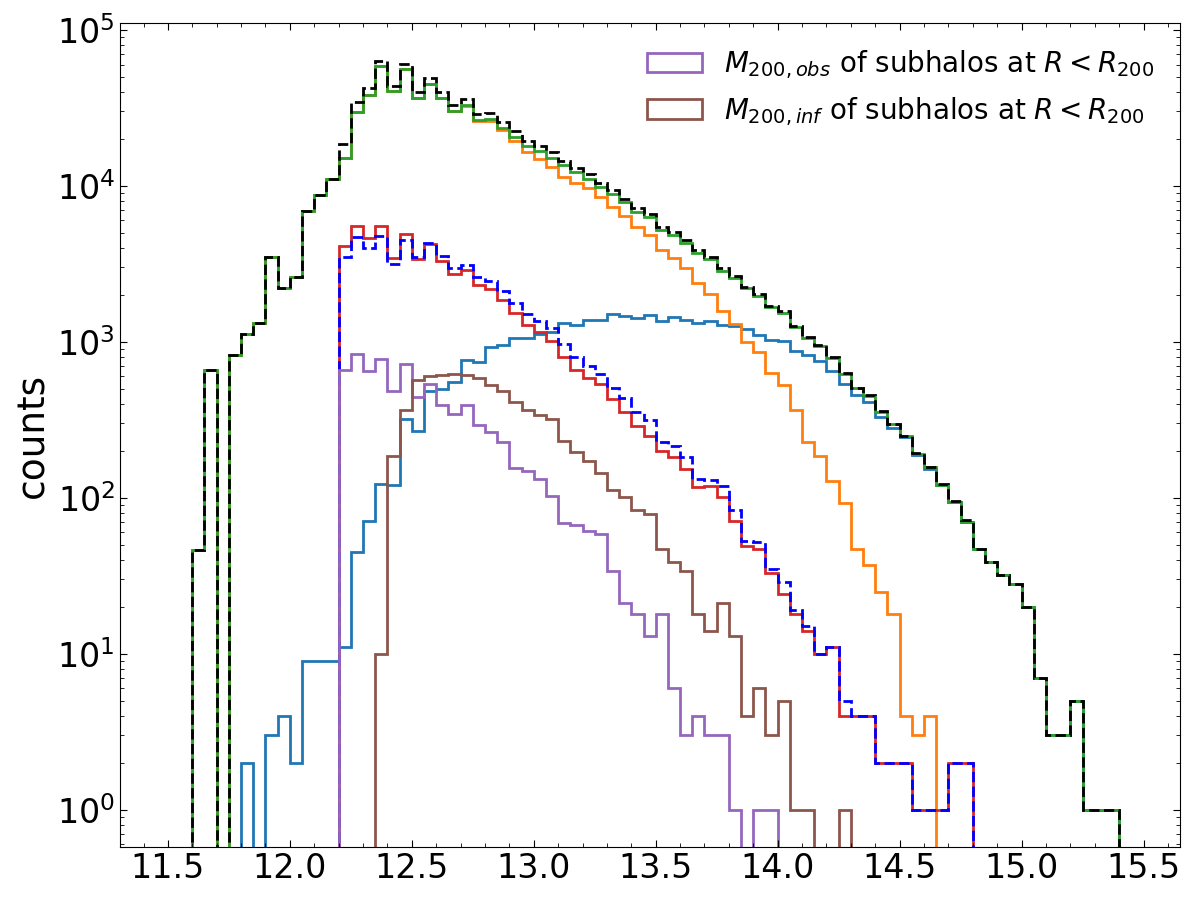}
      \includegraphics[width=0.49\textwidth]{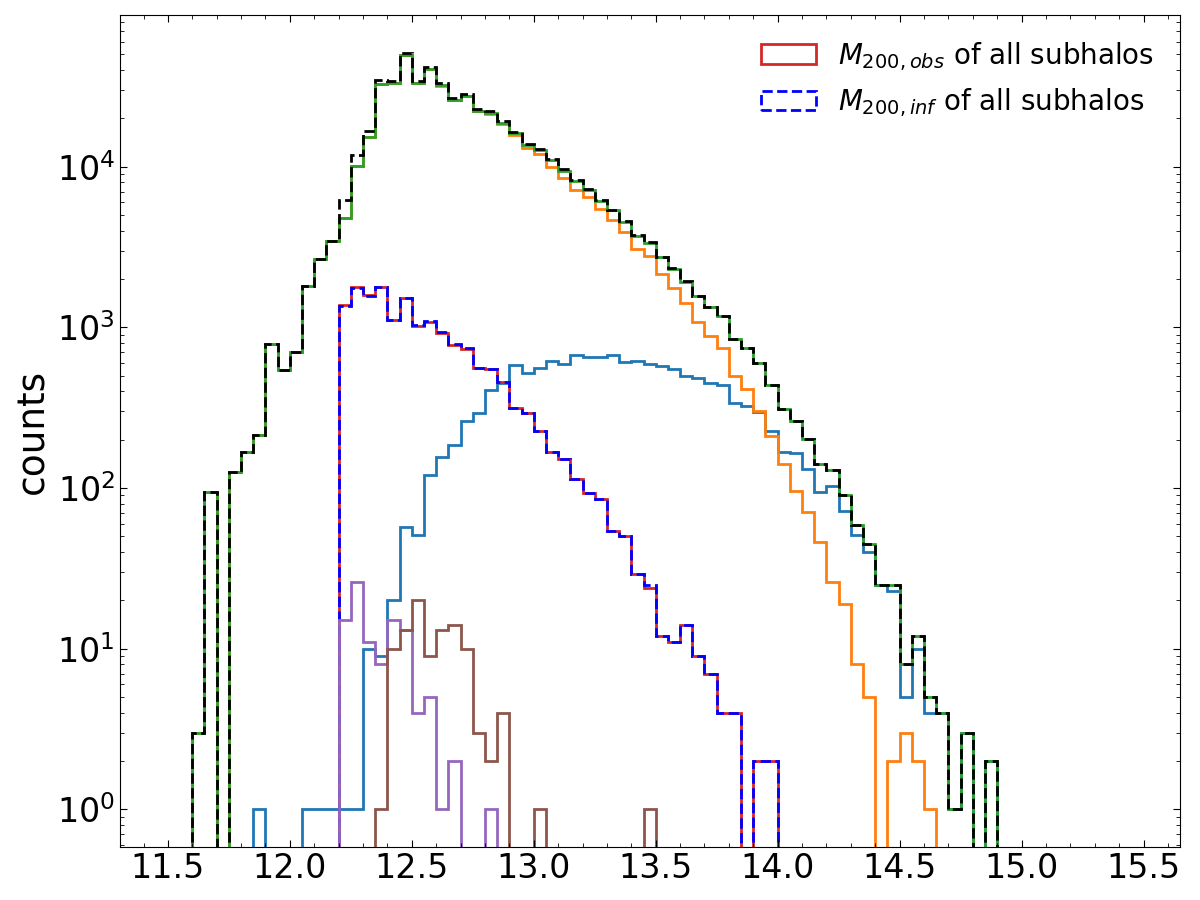}
         \includegraphics[width=0.49\textwidth]{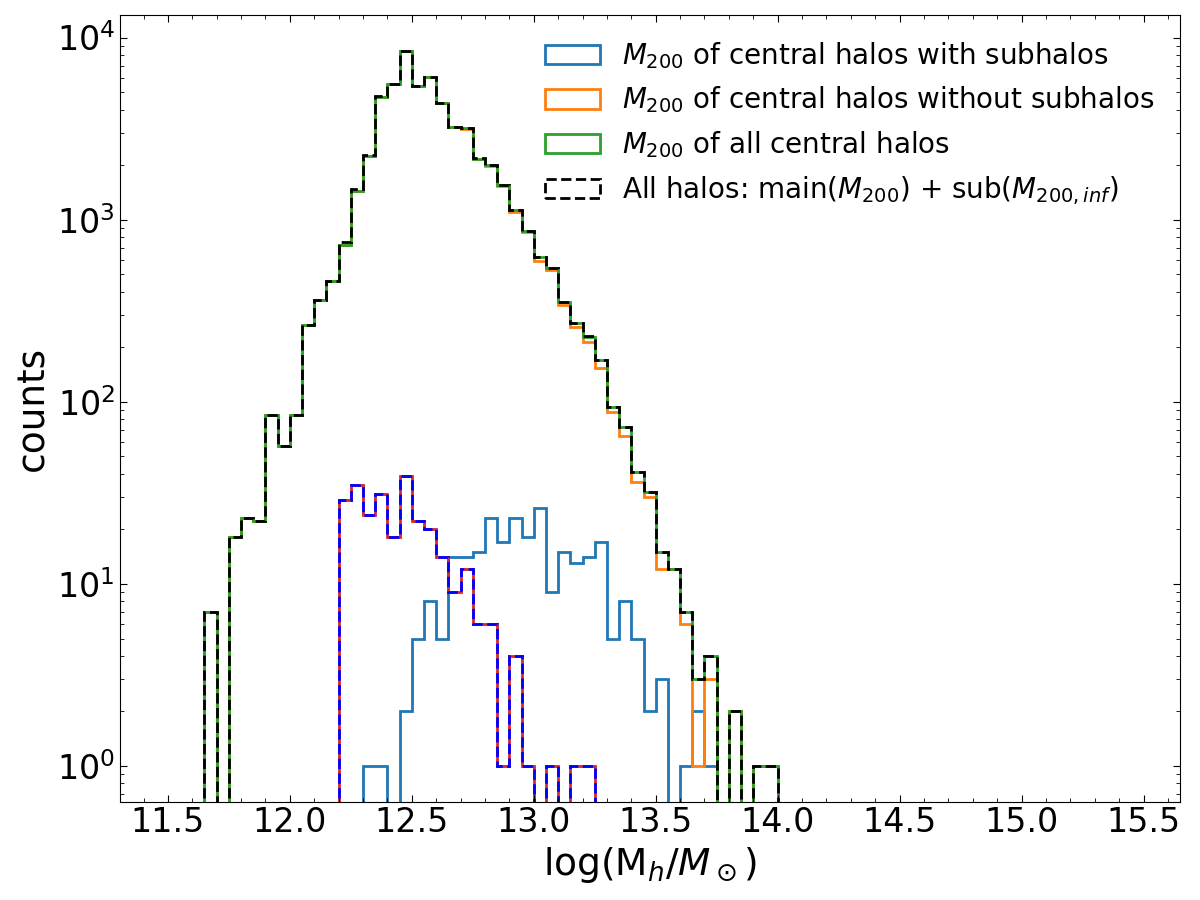}
  \caption{Histograms of halo masses for three comoving boxes located at $z=0$ (top panel), $z=1$ (central panel), and $z=3$ (bottom panel). In all panels, we show the distribution of observed mass (violet) and infall mass (brown) of subhalos that have entered $R_{200}$ of their main halo (i.e., are~located at $R<R_{200,\mathrm{main\,halo}}$). We also show the distribution of observed (red) and infall (blue) masses of all subhalos and the mass distribution of main (or central) halos with subhalos (light blue), the distribution of main halos without subhalos (orange), the total distribution of all main halos (green), and the distribution of all halos (black; i.e., both main and subhalos) considering the infall mass for subhalos.}
\label{infall}
\end{figure}

The collapsed CDM structures were identified in each comoving snapshot by means of a friends-of-friends algorithm (FoF, \citealt{Davis85}) run on the CDM particles with linking length $\lambda = 0.16 \times d$, where $d$ is the mean interparticle separation, retaining only structures with more than $32$ particles, which correspond approximately to $\sim 10^{12.4}\,M_\odot\,h^{-1}$. In addition to these FoF catalogs, the \textsc{Subfind} algorithm \citep{Springel01} was run to identify gravitationally bound structures and to associate standard quantities, such as the radius $R_{200}$, inside of which the mean density $\rho$ is $200$ times the critical density of the universe $\rho_{\rm crit}$ \citep{White01}, and the 
mass $M_{200}$ contained within $R_{200}$, of the main diffuse substructure of each FoF group. 

We used both the comoving boxes and the light cones of the simulation in the analyses.
The comoving boxes have a size of $750\,{\rm Mpc}\,h^{-1}$ on a side and range from $z=0$ to $z=99$ for a total of $33$ stored snapshots. More in detail, the boxes we considered are located at redshifts $z=0.0$, $0.1$, $0.2$, $0.3$, $0.4$, $0.5$, $0.6$, $0.7$, $0.8$, $0.9$, $1.0$, $1.2$, $1.4$, $1.6$, $1.8$, $2.0$, $2.25$, $2.5$, $3.0$, $3.5$, and $4.0$.
The light cones were built using the MapSim routine \citep{Giocoli14} as described in \citet{Giocoli18}, with the particles stored in the snapshots from $z=0$ to $z=4$. 
The particles from different snapshots were distributed according to their comoving distances with respect to the observer, from which the redshift was derived, and according to whether they lay within a defined aperture of the field of view.  
We created $256$ different light-cone realizations, each with an area of $72.18\, \mathrm{deg}^2$, by randomizing the various comoving cosmological boxes. 

We selected only halos with masses $M_{200} \geq M_{\rm min,halo} = 10^{12.5}\,M_\odot$ for the comoving boxes and the light cones because the mass resolution of the simulation is finite. In this way, we ensured that our catalogs are complete, as we show in Fig.~\ref{infall}, which presents the halo mass distribution. 

We also identified subhalos: main (or central) halos contain a population of subhalos, which are the remnants of accreted halos. Because of the resolution of our simulation, subhalos are one order of magnitude less abundant than main halos, as we also show in Fig.~\ref{infall}.
We further note that central halos with subhalos are on average more massive than central halos without subhalos at $z<1$, whose contribution is almost negligible at masses above $M_h\sim 10^{14}\,M_\odot$.
We investigate the characteristics and relative importance of subhalos in the next section.

\subsection{Subhalos and infall mass}

Subhalos are gravitationally bound structures that are smaller than the main halo they belong to, orbiting within the gravitational potential of their main halo. Moreover, when they enter $R_{200,\mathrm{main\,halo}}$, they start to lose mass through various dynamical effects, including dynamical friction, tidal stripping, and close encounters with other subhalos, and they may eventually be completely disrupted (e.g.,~\citealt{Choi07}). Stars are centrally concentrated and more tightly bound than the dark matter, and the stellar mass of a galaxy that is accreted by a larger system is therefore expected to change only slightly, even though most of the dark matter has been stripped off. Therefore, the subhalo mass at the time of observation is not the best tracer for the potential well that shaped the galaxy properties. A better tracer is the subhalo mass at the time it was accreted to the host halo (hereafter, the ``infall mass''). This was first proposed by \citet{Conroy06} and was used in several works (e.g., \citealt{Moster10, Behroozi10,Moster13,Reddick13}).
We followed \citet{Gao04} to parameterize the retained mass fraction of each subhalo as a function of its distance from the host halo center $r$ in units of the radius $R_{200}$ of its host halo,
\begin{equation}\label{eq1}
\frac{M_{\rm obs}}{M_{\rm inf}}=0.65\,\left (\frac{r}{R_{200}}\right)^{2/3}\, ,
\end{equation}
where $M_{\rm inf}$ is the infall mass, $M_{\rm obs}$ is the mass of the subhalo at the moment of observation (i.e.,~after losing mass through interactions).
We also evaluated the infall redshift by calculating the accretion time $t_m$ by inverting the relation presented in \citet{Giocoli08},
\begin{equation}
M_{\rm inf}(t)=M_{\rm obs}\,\exp\left[\frac{t-t_m}{\tau(z)} \right]\, ,
\end{equation}
where $\tau(z)$ describes the redshift dependence of the mass-loss rate.  \citet{Vandenbosh05} proposed the following equation for $\tau(z)$:
\begin{equation}
\tau(z)=\tau_0 \left[\frac{\Delta_V(z)}{\Delta_0} \right]^{-1/2} \left[\frac{H(z)}{H_0} \right]^{-1}\, ,
\end{equation}
with $H(z)$ the Hubble constant at redshift $z$, $\tau_0=2.0\,$Gyr, and $\Delta_V=\rho/\rho_{\rm crit}$ \citep{Bryan98}.

We followed this approach because the merger trees of the simulation were not stored, and therefore the infall mass cannot be calculated directly from the snapshots. However, as shown in \citet{Gao04}, Eq.~\ref{eq1} has been calibrated for subhalos with masses higher than $6 \times 10^{10} h^{-1}M_\odot$ in a $\Lambda$CDM cosmology, and therefore it is suitable for the regime of subhalo masses in our simulation.

In the remainder of this paper, we use the infall mass of the subhalos because as we described above, it is a better tracer of the gravitational potential well at the moment of galaxy formation. It is therefore expected to provide a better link to the stellar mass than the observed mass. In Fig.~\ref{infall} we show the histograms representing the mass distribution of subhalos that have lost mass (i.e.,~those located at $r<R_{200}$ at the time of observation) for three comoving boxes, and for the mass at the time of observation and the infall mass. For subhalos at $r<R_{200}$  the median mass loss at all redshifts is $\delta log(M_h)=0.25\, \mathrm{dex}$, with a maximum mass loss of $\delta log(M_h)_{\rm max}=1.26\, \mathrm{dex}$ (this only occurs in a few cases).
Fig.~\ref{infall} shows, however, that because of the mass resolution of the simulation we adopted, the total contribution of subhalos to the density of the total population of halos is almost negligible.

It might be argued that galaxies can also lose or acquire stellar mass through gravitational interaction between different galaxies in different halos (or subhalos), or through simple star formation activity, since the time of the infall. The median time elapsed between the infall of the subhalo into the main halo and the observation time is $\delta(t)=0.6$\,Gyr. The stellar mass loss through gravitational interactions can be modeled through simulations of galaxy formation and evolution. As an example, \citet{Kimm11} adopted a semianalytic approach and tried to model the mass loss of satellite galaxies at the moment they enter into their main halos: the authors found that the majority of baryonic mass loss is in the form of hot and cold gas that is present in the disk or halo of the galaxy. The stellar mass loss is not trivial to model, and quantitative values were not provided. On the other hand, galaxies continue to form stars with a rate that depends on their star formation history, gas reservoirs, initial mass function (IMF), and several other parameters.  Using the Illustris hydrodynamic simulation, \citet{Niemiec19} found that subhalos can lose a large portion of their dark matter at accretion, but continue to form stars, which increases the stellar mass up to $\sim 6$\% after $1$\,Gyr from the accretion event.
Considering the median timescale of $0.6$\,Gyr, the uncertainties in treating the processes of stellar mass loss and gain, and the numerous variables at play, we decided to take  the stellar mass changes after the infall into the main halos not into account.
In the following, unless specified otherwise, the halo mass $M_h$  represents
\begin{equation}
M_h=\begin{cases} M_{200},\, \mathrm{for\,\, main\,\, halos}\\ M_{\rm inf},\, \mathrm{for \,\, subhalos}
\end{cases}.
\end{equation}


\subsection{Halo mass function}

\begin{figure*}
   \centering
   \includegraphics[width=0.86\textwidth]{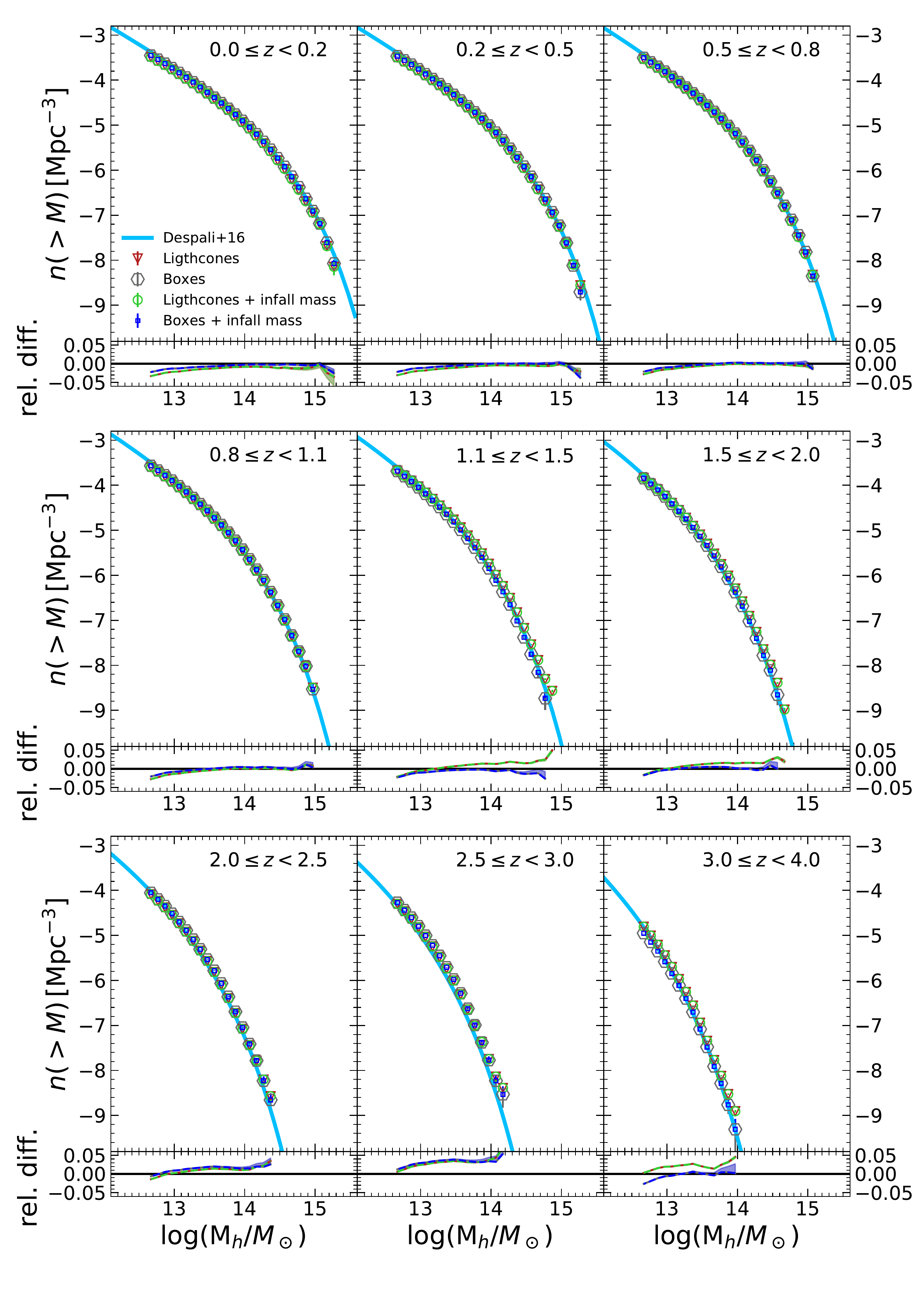}
  \caption{Cumulative halo mass functions in $\text{nine}$ different redshift bins. The cyan line represents the \citet{Despali16} halo mass function calculated at the center of the redshift bin. The colored points represent the halo mass function of the $\Lambda $CDM {\small DUSTGRAIN-}{\it pathfinder} simulation: red triangles are the halo CMF evaluated on the ligth cones without taking the infall mass for the subhalos into account, green circles make use of the infall mass, gray hexagons represent the halo CMF using comoving boxes without using the infall mass for the subhalos, and blue squares are derived from simulation boxes taking the infall mass into account. The bottom panels represent the relative difference of the CMF of the simulation with respect to the \citet{Despali16} mass function.}
\label{allHCMF}
\end{figure*}

The mass function is defined as the number density of objects (i.e.,~halos and galaxies) per unit comoving volume in bins of (halo or stellar) mass and redshift.  In order to extend the SHMR not only at any redshift, but also in the widest possible mass range (because the simulation we used is limited to halos with $\log (M_h/M_\odot)\geq 12.5$), we compared the halo mass functions (HMFs) of the simulation to theoretical parameterizations. 
In Fig.~\ref{allHCMF} we show a comparison of the cumulative halo mass functions (CMF) of the $\Lambda $CDM {\small DUSTGRAIN-}{\it pathfinder} simulation with those reported by \citet{Despali16}, who measured the HMF using a suite of six N-body cosmological simulations with different volumes and resolutions (see \citealt{Despali16} for details). We computed the \citet{Despali16} cumulative HMFs for halos whose mass is defined as $200\,\rho_{crit}$ in order to match the halo mass definition of the simulation we used. 
In particular, we show the CMF computed from light cones without taking the infall mass for the subhalos into account, and the one using the infall mass. Using the light cones, we were able to precisely select the redshift intervals in which we calculated the CMF. The comoving boxes, instead, are located at fixed redshifts. To calculate the HMF, we therefore considered all the boxes located in the same redshift bin.  We also evaluated the halo CMF for boxes with or without the infall mass for the subhalos. 
We estimate the relative difference of the CMF of the simulation with respect to the \citet{Despali16} mass function in the bottom panels of Fig.~\ref{allHCMF}, showing that the differences between the HMFs of the simulation (on comoving boxes and light cones) and the theoretical HMFs are minimal: the maximum difference is only a few percent of the value of the \citet{Despali16}. Fig.~\ref{allHCMF} also clearly shows that even if the median difference between the infall mass and observed mass is $0.25$\,dex, the halo mass function does not change signficantly, regardless of the infall mass for subhalos. When we consider the mass of subhalos at the infall or the mass at observation, the differences in mass functions are lower than $1\%$. This is reasonable because only a few subhalos have already entered $R_{200}$, and have therefore lost mass. This is because the simulation is limited to very massive halos ($\log(M_h/M_\odot)\geq 12.5$).  Their
total influence on the HMF at high halo masses and on the derived SHMR is
therefore very limited because the fraction of subhalos with respect to main halos is small ($\sim$ 10\% at $\log(M_h/M_\odot)= 12.5$ and $z=0$, consistent with the results by \citealt{Rodriguez17} and \citealt{Despali17}). This is also visible in the histograms of Fig.~\ref{infall}, which clearly show that the effect of subhalos is negligible compared to the total halo distribution, regardless of the infall mass.

In order to derive the SHMR relation on a wider mass range than allowed by the HMFs of $\Lambda $CDM {\small DUSTGRAIN-}{\it pathfinder} simulations, we adopted the \citet{Despali16} HMF: in this way, we were able to compute the SHMR from the lowest observed stellar masses ($M_*\sim 10^8\,M_\odot$, $M_h\sim 10^{10.5}\,M_\odot$) to the highest ($M_*\sim10^{12}\,M_\odot$, $M_h\sim10^{15}\,M_\odot$) from $z=0$ to $z=4$. Finally, we stress that in order to build a realistic mock catalog of galaxies, however, the derived SHMR relation needs to be applied to the infall mass of subhalos to derive the corresponding stellar mass.

\section{Stellar mass functions}\label{observedSMF}

The SMF is a powerful tool for statistically describing the distribution of stellar mass in galaxies, the galaxy mass assembly over cosmic time, and the evolution of the galaxy population with redshift. The SMF of galaxies has been extensively studied over the past years out to $z \sim 4 - 5$ (e.g., \citealt{Fontana06,Pozzetti07,Stark09,Pozzetti10,Ilbert13,Muzzin13,Grazian15,Song16,Davidzon17}). Traditionally, the SMF is modeled with a Schechter function \citep{Schechter76}, but for certain galaxy populations, a double Schechter function provides a better fit to observations (e.g., \citealt{Baldry08,Pozzetti10}). 
For all the SMFs we used in this work that we describe in the following sections, we rescaled the data points to our cosmology (i.e., to Planck15 values) and to a Chabrier IMF.

\subsection{Data at $z\sim 0$}\label{baldry}

We used the SMF at $z\sim 0$ derived by \citet{Baldry08} for galaxies in the New York University Value-Added Galaxy Catalog (NYU-VAGC) sample of galaxies \citep{Blanton05} derived from the Sloan Digital Sky Survey (SDSS). The data were obtained from the SDSS catalog and cover cosmological redshifts from $0.0033$ to $0.05,$ where the redshifts were corrected for peculiar velocities using a local Hubble-flow model \citep{Willick97}. The NYU-VAGC low-$z$ galaxy sample was used to recompute the SMF down to low masses. The Data Release 4 (DR4) version of the NYU-VAGC low-$z$ sample includes data for $49\,968$ galaxies.
\citet{Baldry08} determined the stellar masses by fitting Petrosian $ugriz$ magnitudes of each galaxy to the observed frame
using the NYU-VAGC magnitudes and PEGASE SPS models \citep{Fioc97} and a variety of different extinction laws (from a Small Magellanic Cloud screen law to a $\lambda^{-0.7}$ power law). The authors also showed that their results obtained with PEGASE models or BC03 models \citep{B&C03} are consistent with each other. When no stellar mass was available for a galaxy, the stellar mass was determined using a color–$M/L$ relation calibrated to the particular set of stellar masses. 
Moreover, these data were matched to stellar masses estimated by \citet{Kauffmann03}, \citet{Gallazzi05}, and \citet{Panter07}. 
Finally, \citet{Baldry08} averaged between the different stellar mass estimates by applying a weight depending on the normalized number density as a function of redshift for all four mass estimates, and by recomputing the SMF of the sample. They obtained an accurate SMF for $z\sim0$ galaxies. In this work, we use the best-fit Schechter function of the SMF evaluated on \citet{Baldry08} datapoints.

\citet{Bernardi13} discouraged the use of Petrosian magnitudes to derive masses and luminosities in SDSS, advocating possible problems in the resulting mass-to-light ratios; this could affect the massive end of the SMFs, which might be underestimated. We note that the expected effect only concerns the brightest objects at very low redshift. The flux loss on datasets at higher redshifts, described in the next section, has not yet been established.

\begin{figure}[ht]
   \centering
   \includegraphics[width=9cm]{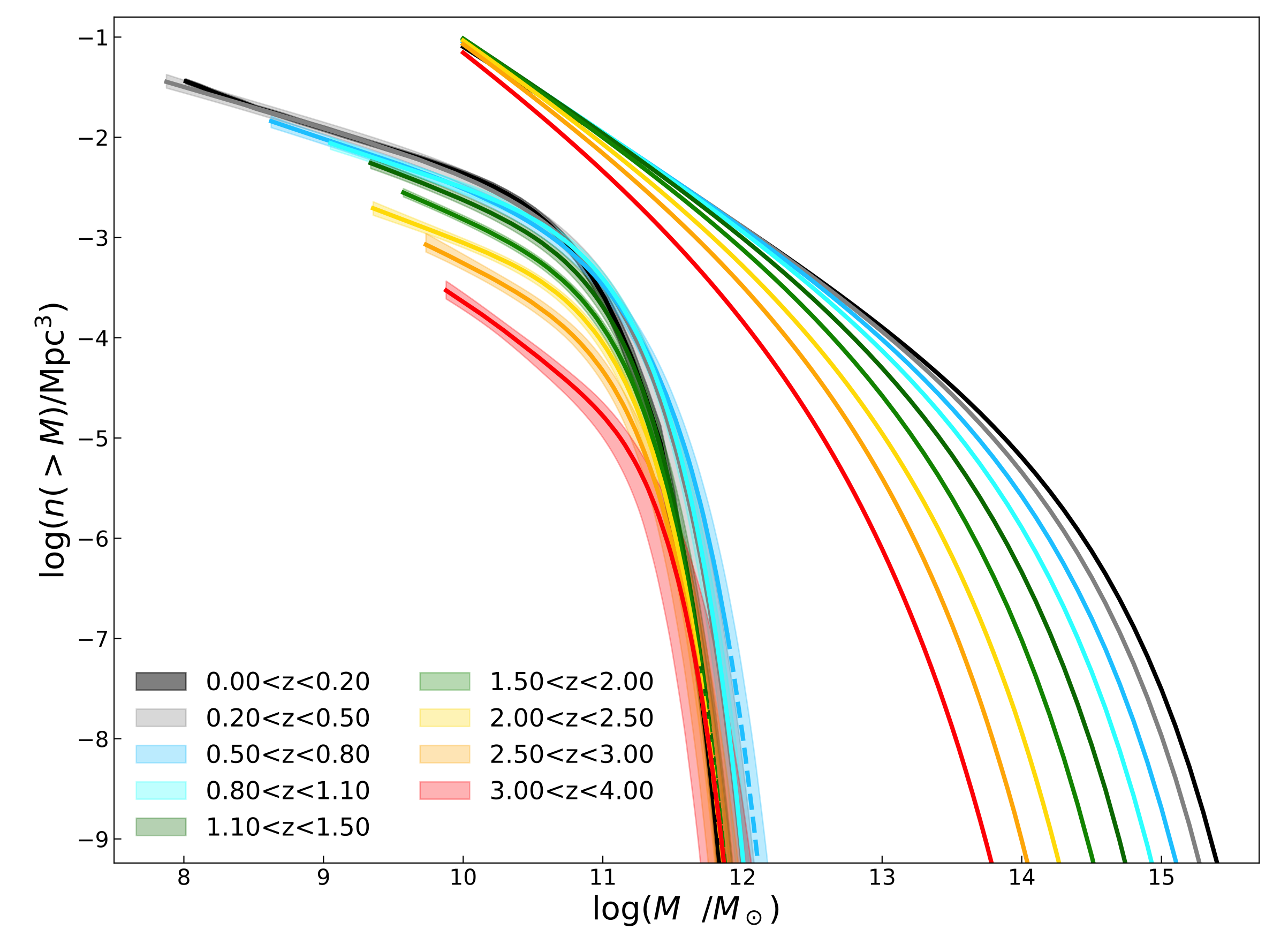}
  \caption{The CMFs used in the (sub-) halo abundance matching in different redshift bins. On the left, solid and dashed lines with shaded area represent the SMF with $1\sigma$ uncertainty (best fit to $1/V_{\rm max}$ points; with the exception of the redshift bin $0.0<z<0.2$, the best fits have been corrected for the Eddington bias) cumulated as described in Sect.~\ref{SHAM}. Solid lines represent the SMF up to the last $1/V_{\rm max}$ point, and dashed lines show this at higher masses.  On the right, the solid lines are the \citet{Despali16} CMFs we used to evaluate the SHMR.}
              \label{CMFs}%
\end{figure}
\subsection{Data at $z>0$}\label{Ilbert}

For $z>0$ data, we adopted the \citet{Ilbert13} SMFs, which were estimated in the COSMOS field using one of the largest dataset. COSMOS \citep{Scoville07} is one of the best available fields to derive the SMF because the area is quite large ($\sim$ 2 deg$^2$) and many deep ($I_{\rm AB}\sim 26.5$) multiwavelength data are available in more than $35$ bands.
\citet{Ilbert13} selected the sample using near-infrared data from the UltraVISTA DR1 data release \citep{McCracken12}. The sample was built by restricting the analysis to objects with $K_s<24$ and to sources in regions with good image quality, totalling an area of $1.52\, \mathrm{deg}^2$. The photometric redshifts and stellar masses were derived with high precision by fitting the spectral energy distribution (SED) with \textit{LePhare} code \citep{Arnouts02,Ilbert06} and Bruzual and Charlot stellar population synthesis models \citep{B&C03}. To compute photometric redshifts, a variety of extinction laws were considered \citep{Prevot84, Calzetti00} and a modified version of the Calzetti law that includes a bump at $2175\AA$, while in the computation of stellar masses they only considered the \citet{Calzetti00} extinction law. Galaxies with masses as low as $M \sim 10^{10}M_\odot$ are detected up to $z = 4$. More generally, a minimum mass $M_{\rm limit}$ was defined as the $90\%$ completeness limit and was used as the lower boundary in the evaluation of the SMF. 

\citet{Ilbert13} estimated the SMF in eight redshift bins from $z=0.2$ to $z=4.0$ using different methods to determine the possible biases. Here we used the best-fit Schechter function on the binned $1/V_{\rm max}$ \citep{Schmidt68} points and relative uncertainties, computed by adding the errors due to galaxy cosmic variance in quadrature to the template-fitting procedure and the Poissonian errors. 

To derive an SHMR that is as general as possible and does not depend on the quality of the observed dataset that is used to derive the SMF, we used the \textit{\textup{intrinsic}} SMF, derived by \citet{Ilbert13}, which accounts for the uncertainties in the stellar mass estimate. 
Because the galaxy density exponentially decreases toward massive galaxies, errors in the stellar mass scatter more galaxies toward the massive end than in the opposite direction \citep{Eddington}. This  biases the estimate of the high-mass end \citep{Kitzbichler07,Caputi11}. The detailed procedure for correcting for this bias is described in \citet{Ilbert13}, and permits determining the intrinsic SMF: the stellar mass uncertainties are well characterized by the product of a Lorentzian distribution $L(x)=\frac{\tau}{2\pi}\frac{1}{(\tau/2)^2 +x^2}$ with $\tau=0.04(1+z)$ and a Gaussian distribution $G$ with $\sigma = 0.5$. The observed SMF is the convolution of the intrinsic SMF, parameterized with a double Schechter function $\phi$, and the stellar mass uncertainties: $\phi_{\rm convolved}=\phi\,(L\times G)$. By estimating $\phi_{\rm convolved}$ that fits the $1/V_{\rm max}$ datapoints, it is therefore possible to determine the intrinsic $\phi$.

For $z\sim 0$ data, we simply used the best-fit Schechter function from the \citet{Baldry08} data without the deconvolution to determine the intrinsic $\phi$: this effect would act in the opposite direction of the flux underestimation suggested by \citet{Bernardi13}. Because we lack a precise estimate of the two effects, we preferred to use face value data. Moreover, the possible effect on the SHMR is smoothed out when the model is derived, as explained in Sect.~\ref{sectSHMR}.

To summarize, we adopted the intrinsic SMFs at $z>0$ as the best fit to evaluate the SHMR. The intrinisc SMFs were statistically corrected for errors due to stellar mass and photo-$z$ uncertainties, and take Poissonian and cosmic variance errors into account. We emphasize here that COSMOS is still one of the best datasets that can be used to derive the  SMFs because of the high-precision photometric redshifts and stellar masses, and because it allows working with a large and homogeneous statistical sample over a wide range of redshifts.

\section{The stellar-to-halo mass relation}
\label{SHMR}

In this section we derive the relationship that connects the stellar mass of a galaxy to the mass of its dark matter halo. We also detail the procedure to calculate the empirical model and its redshift evolution.

\subsection{Subhalo abundance matching technique}
\label{SHAM}

In order to link the stellar mass of a galaxy $M_*$ to the mass of its dark matter halo $M_h$, we derived the SHMR using the subhalo abundance matching technique. This method relies on the assumption that the halo mass monotonically relates to a galaxy property, that is, to the stellar mass. 
We assumed that every main (sub-) halo contains exactly one central (satellite) galaxy and that each halo is populated with a galaxy.
In other words, we assumed a deterministic one-to-one relation between halo and stellar mass. To uniquely associate one halo to one galaxy, we built the cumulative version of the halo mass function $n(M_h)$ and of the galaxy SMF $\Phi(M_*)$. The latter was built by simply summing the number counts of objects with stellar masses greater than the considered one, $M_i$:
\begin{equation}
n(>M_i)=\int^{\infty}_{M_{i}} \Phi(M')dM'\, .
\end{equation}

In Fig.~\ref{CMFs} we display the stellar and halo CMFs, adopting for the latter the parameterization proposed by \citet{Despali16}.
As discussed in the previous section, the \citet{Despali16} halo CMF represents a good approximation of the (sub-) halo mass function from our simulation, even when the correction for the infall mass is taken into account. For this reason, we adopted it to evaluate the SHMR on the largest possible mass range.
Finally, from the direct comparison of the cumulative stellar and halo mass functions, we derived the ratio between $M_*$ and $M_h$ at each fixed number density. We repeated this operation in each redshift bin from $z=0$ to $z=4$.

When available, we adopted the \textit{\textup{intrinsic}} SMFs, which were deconvolved for all the errors associated with the stellar mass calculation, as described in Sects.~\ref{baldry} and \ref{Ilbert}. In this way, we evaluated the \textit{\textup{intrinsic}} SHMR. However, when realistic mock catalogs are built, an error in stellar mass at fixed halo mass needs to be applied to reproduce the observational effects in the estimate of the stellar mass from observed photometry. In other words, to perform a comparison with the observed and not the intrinsic SMFs, a scatter  needs to be applied.  We added this error as an {\it \textup{observational}} scatter ($\sigma_{obs}$) from a log-normal distribution and explore its effects on the SHMR and the derived SMFs in Sect.~\ref{scatter}.

However, we expect that in nature, two halos of the same mass may harbor galaxies with different stellar masses and vice versa because they can have different merger histories, spin parameters, and concentrations. This can be taken into account by introducing an additional {\it \textup{relative}} scatter ($\sigma_R$) in the intrinsic SHMR relation.
In the literature, several approaches have been followed: in general, the scatter is drawn from a log-normal distribution with a variance that depends on the assumptions of the different analyses. Some works assume a scatter with constant variance at all masses and redshifts (e.g., $\sigma_R=0.15$\,dex for \citealt{Moster10,Moster13}). Other works,  such as \citet{Behroozi10,Behroozi13,Legrand19}, instead keep the variance as an additional free parameter and fit its value, keeping it constant with halo mass but not with redshift. As an example, \citet{Legrand19} found values that range from $0.14$ at $z\sim 0.35$ to $0.46$ at $z\sim 5$. Using X-ray observations on clusters, \citet{Erfanianfar19} found a mean scatter of $0.21$ and $0.25$\,dex for the stellar mass of the brightest cluster galaxies in a given halo mass at low ($0.1<z<0.3$) and high ($0.3<z<0.65$) redshifts, respectively.
In addition, recent results from hydrodynamic simulations (e.g., Eagle, see \citealt{Schaye15}) have shown that this scatter depends on halo masses \citep{Matthee17}, ranging from $0.25$\,dex at $M_h=10^{11}\, M_\odot$ to $0.12$ at $M_h=10^{13}\, M_\odot$. \citet{Matthee17} also found a weak trend for halo masses above $M_h=10^{12}\, M_\odot$. Recent works, such as \citet{Moster18}, self-consistently introduced scatter in the SHMR by taking the full formation history of halos into account. They found a scatter that depends both on halo masses and redshifts.

Therefore we also evaluated the SHMR in the presence of  {this {\it \textup{relative}}} scatter {in stellar mass at fixed halo mass.} 
In particular, we followed the approach detailed in \citet{Moster10}. 
We first evaluated the model SMF that is to be convolved with this scatter to reproduce the observed intrinsic SMF. 
We then derived the SHMR through its direct comparison with the HMFs.
We fixed the standard deviation ($\sigma_R$) of a log-normal distribution to $0.2$ dex at all halo masses and redshifts (as done in \citealt{Moster10,Moster13}). 
Similar values have been  presented in several observational works (e.g., \citealt{Erfanianfar19}), in various empirical modeling or abundance matching works \citep[e.g.,][]{Moster18, Legrand19}, and hydrodynamic simulations \citep[e.g.,][]{Matthee17}. In the remainder of this work we present results that were calculated either with or without the {\it \textup{relative}}  scatter. However, given the current uncertainties of the value of this scatter and its unknown dependence on redshift and mass, we prefer to present results as {\it \textup{reference}} for the case calculated without the scatter. We show, however, that qualitative results are preserved when we include this {\it \textup{relative}} scatter, while the differences quantitatively depend on the exact value of the scatter that is introduced.

\subsection{Parameterizing the SHMR}
\label{evolutionz0}

In order to parameterize the SHMR that we derived in each redshift bin, we adopted the simple double power-law function proposed by \citet{Moster10},
\begin{equation}\label{mosterrel}
\frac{M_*}{M_h}(z)=2\,A(z)\, \left[\,\left(\frac{M_h}{M_A(z)}\right)^{-\beta(z)}+\left(\frac{M_h}{M_A(z)}\right)^{\gamma(z)}\right]^{-1}\, ,
\end{equation}
where $A$ is the normalization of the SHMR at the characteristic halo mass $M_A$, while $\beta$ and $\gamma$ describe the slopes of the relation at low- and high-mass ends, respectively.

\subsection{Evolution with redshift of SHMR parameters}
\label{evolutionzbig}

In this section we define a redshift dependence of the SHMR parameterization in order to construct an empirical model that describes its evolution with cosmic time.
Studying the evolution of the SHMR helps place constraints on the processes related to galaxy formation and evolution. Furthermore, with this information, we can populate the N-body simulation light cones with galaxies at different redshifts using the appropriate redshift-dependent SHMR.
Following \citet{Moster10}, to determine the SHMR at any redshift, we adopted a redshift dependence for the parameters of Eq.~\ref{mosterrel}.
\begin{equation}\label{1}
\log M_A(z)=(\log M_A)_{z=0}+ z\cdot \mu =B+ z\cdot \mu \, ,
\end{equation}

\begin{equation}\label{2}
A(z)=\left(\frac{M_*}{M_h}\right)_{z=0}\cdot (1+z)^\nu=C\cdot (1+z)^\nu \, ,
\end{equation}

\begin{equation}\label{3}
\gamma(z)=\gamma_0\cdot(1+z)^{\eta}=D\cdot (1+z)^\eta\, ,
\end{equation}

\begin{equation}\label{4}
\beta(z)=F\cdot z +E\, ,
\end{equation}

\section{Results}
\label{sectSHMR}

\subsection{Relation of the stellar to halo mass}
\label{obsSHMR}

\begin{figure}[ht]
   \centering
   \includegraphics[width=9cm]{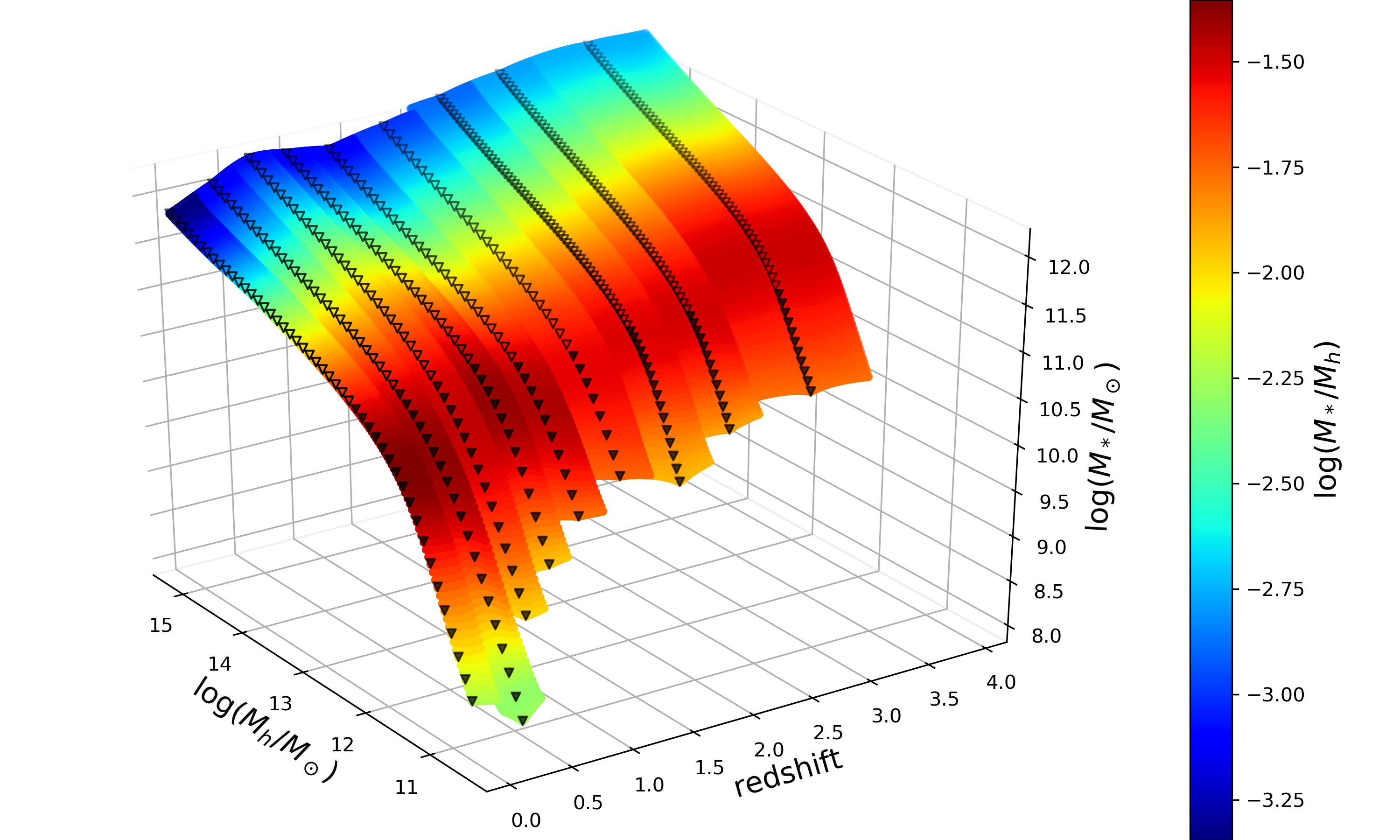}
  \caption{The SHMR without scatter derived in the $\log(M_*)\,-\,\log(M_h)\, $ redshift plane, color-coded with $\log(M_*/M_h)$. Points represent the observed relation derived from the CMFs. Open points show the halo mass range of the simulation (i.e., $\log(M_h/M_\odot)\geq 12.5$), and open and solid points indicate the halo mass range we used to derive the SHMR with the \citet{Despali16} HMF. The colored layer is a linear interpolation between observed data points and is color-coded for the value of $\log(M_*/M_h)$. The lower limit in mass is given by the observed SMFs.}
              \label{SHMR3D}
\end{figure}

\begin{figure*}[h]
   \centering
   \includegraphics[width=16cm]{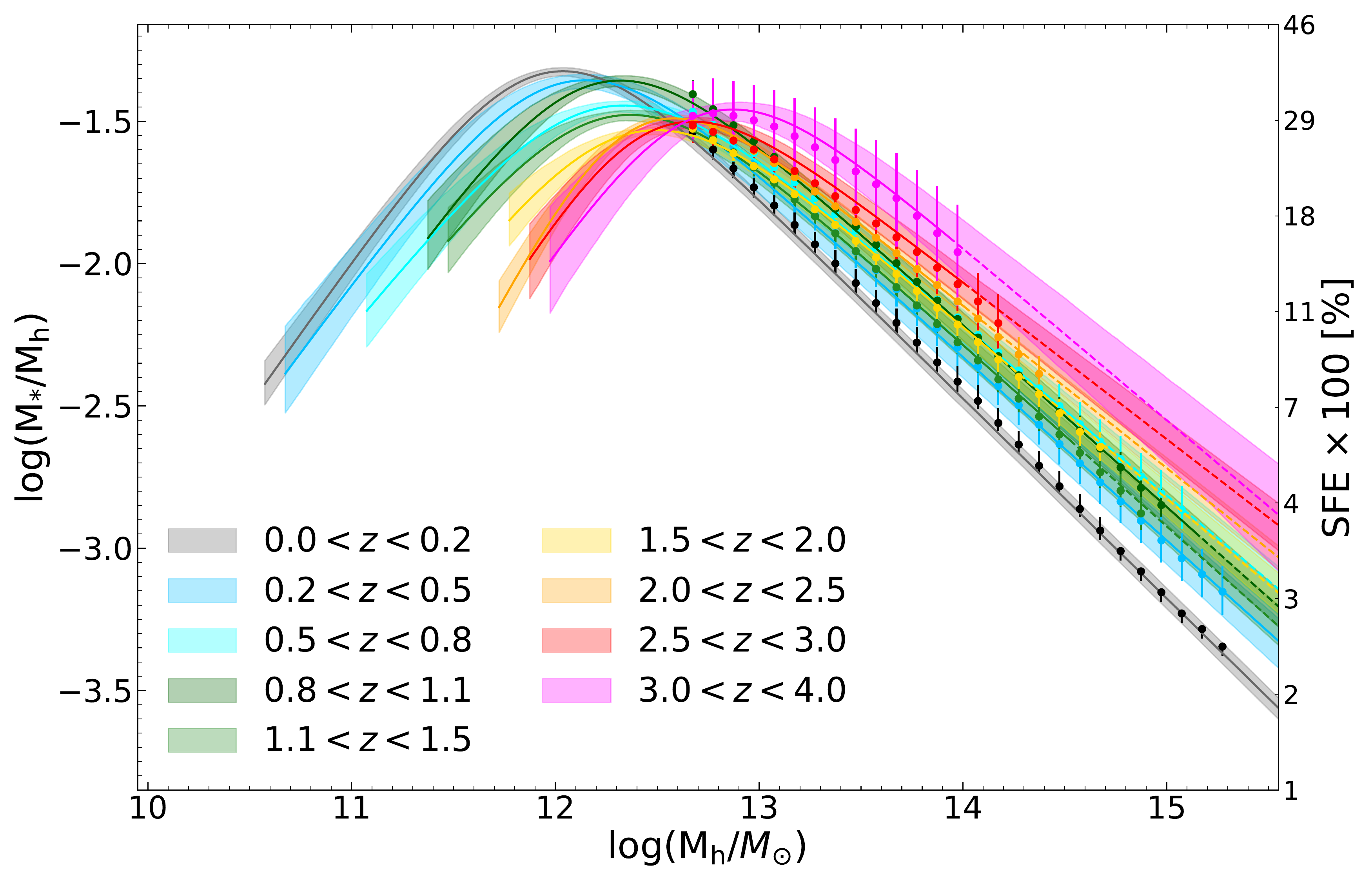}
  \caption{The SHMR for our reference case without scatter derived in the $\log(M_*/M_h)\,-\,\log(M_h)$ plane. Points with error bars represent the observed relation obtained from the CMFs of the light cones (i.e., they represent the halo mass range of the $\Lambda $CDM {\small DUSTGRAIN-}{\it pathfinder} simulation), and the lines and corresponding shaded area represent the relation derived using the \citet{Despali16} mass function and the 1$\sigma$ error bar. The solid line identifies the observed mass range of galaxies, and the dashed line represents the extrapolation using the best-fit Schechter function. The right-hand side y-axis is labeled with the SFE expressed in percentage (defined as $M_*/M_h\,f_b^{-1}$)}
              \label{SHMR_obsfit}%
\end{figure*}

Figures \ref{SHMR3D} and \ref{SHMR_obsfit} show the relation between stellar and halo masses at different redshifts, respectively, and the corresponding ratio (SHMR) for our samples in the redshift range $0 < z < 4$. Figure~\ref{SHMR3D} shows our results in the 3D  space $\log(M_*)\,-\,\log(M_h)\,-$\,redshift plane, color-coded for $\log(M_*/M_h)$.
We show the measurement of the direct comparison of the stellar with halo CMFs{ for our {\it \textup{reference}} case, } that is,  without the {\it \textup{relative}}  scatter. 
For each of the nine redshift bins, we performed a fit using Eq.~\ref{mosterrel} and the software package \textit{emcee}, which is a purely python implementation of the Monte Carlo Markov chain (MCMC) method \citep{Foreman12}. 
This algorithm allows sampling the posterior distribution for the four free parameters (i.e., $M_A$, $A$, $\gamma,$ and $\beta$). We used $200$ walkers (each performing $1000$ steps) with a different starting point each randomly selected from a Gaussian distribution around the original starting prior (chosen from the results presented in \citealt{Moster10}). The first half of the steps were discarded as a burn-in phase. The convergence of the fits was assessed through the Gelman-Rubin diagnostic \citep{Gelman92}. This test compares the variance within one chain of the MCMC with the variance between chains. The two variances are combined in a weighted sum to obtain an estimate of the variance of a parameter. The square root of the ratio of this estimated variance within the chain variance is called the potential scale reduction $\hat{R}$, and for a well-converged chain it should approach $1$. Values higher than $1.1$ indicate that the chains have not yet fully converged \citep{Gelman92}. In the case of the fit in the nine redshift bins, we always find a value of $\hat{R}\leq 1.06$ (ranging from 1.01 to 1.06). 
The best-fit parameters are listed in Table~\ref{indfit}. Values and errors reported in this table were evaluated using a one-to-one relation without any scatter.

\begin{figure}[ht]
   \centering
   \includegraphics[width=9cm]{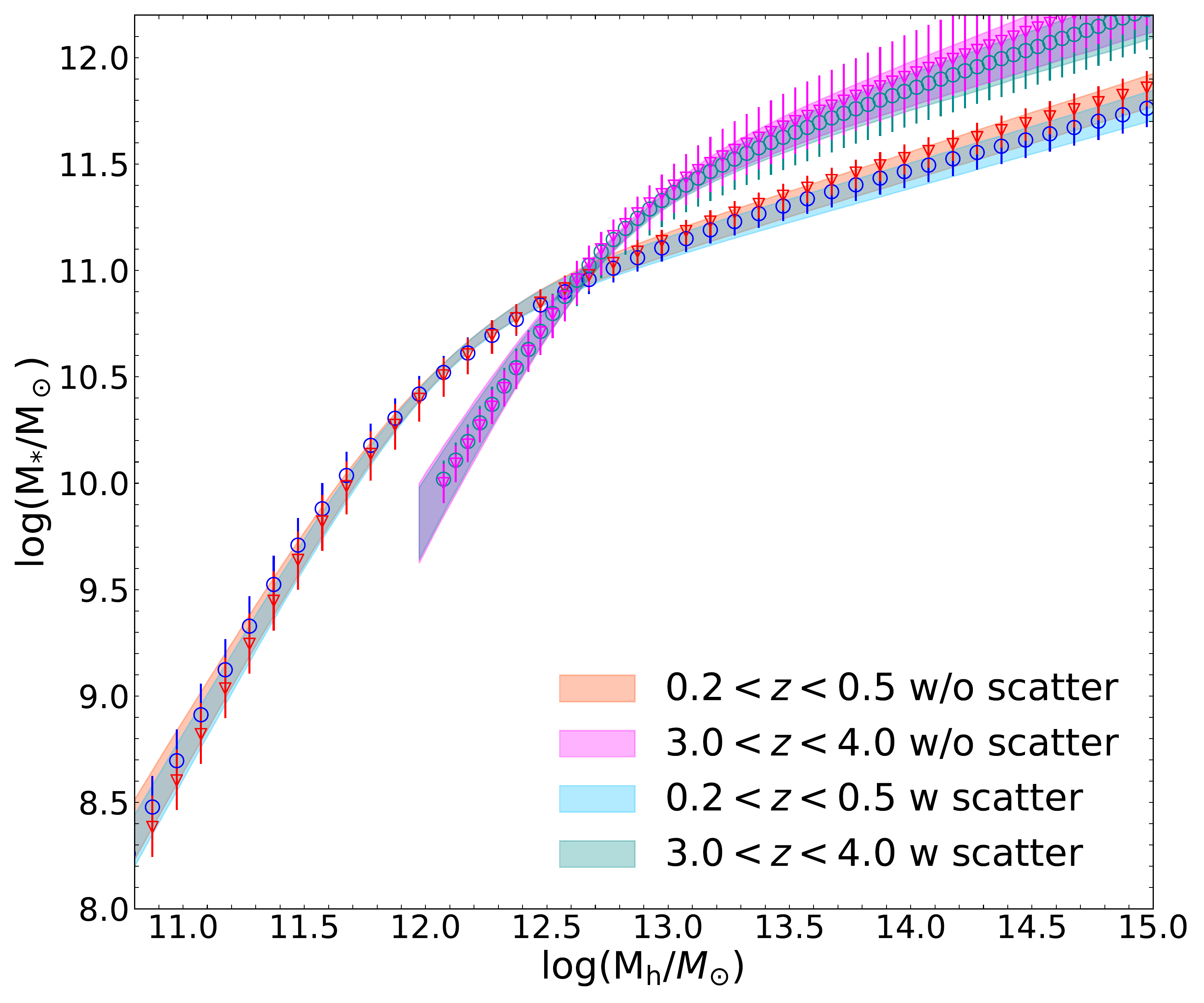}
  \caption{ Comparison of the SHMR evaluated with or without the {\it \textup{relative}} scatter in two different redshift bins. We show the points that represent the direct comparison of the CMFs and the $1\sigma$ confidence region of the best-fit using Eq.\ref{mosterrel} with shaded regions.}
  
              \label{SHMR_comp}
\end{figure}

In Fig.~\ref{SHMR_comp} we show a comparison of the SHMR derived with and without the {\it \textup{relative}} scatter ($\sigma_R$). The points show that scatter mainly affects the massive end of the SHMR (because it mostly affects the massive-end slope of the SMF) but also more weakly affects the low-mass end. At high halo masses (i.e., $\log(M_h/M_\odot)\gtrsim 13$), the SHMR without scatter predicts higher stellar masses at fixed halo mass. At low halo masses (i.e., $\log(M_h/M_\odot)\lesssim 12$), the tendency is reversed. This is due to the convolution we applied to the model SMF when we compared it to the observed SMF to evaluate the SHMR because the integrated number density needs to be conserved. As in the case without scatter, we performed a fit using Eq.\ref{mosterrel}. 
 
In Fig.~\ref{SHMR_comp} we show the $1\sigma$ confidence regions.
The introduction of a{ \it \textup{relative}} scatter mainly affects parameter $\gamma$, which controls the massive end slope but also influences all other parameters because they are all correlated. We also list the best-fit values including{ \it \textup{relative}} scatter in Table.\ref{indfit_scat}.
 We note that the effect of the scatter is always on the order of a few percent on the SHMR, even if its effect is systematic and depends on the exact value of the scatter introduced.

Figure~\ref{SHMR_obsfit} shows our results for our {\it \textup{reference}} case, that is, without {\it \textup{relative}} scatter, in a $\log(M_*/M_h)\,-\,\log(M_h)$ plane in the different redshift bins used by \citet{Ilbert13} to compute the SMF. In this case, we show the points that were computed using the halo CMF of light cones, which take the subhalo infall masses into account, and the best fit of the relation evaluated using the \citet{Despali16} CMF that was fit with Eq.~\ref{mosterrel}, along with the corresponding $1\sigma$ uncertainties. The best-fit values and the uncertainties were computed using the 50th, 16th, and  84th percentiles of the posterior distributions of the free parameters of the fit. They are listed in Table~\ref{indfit}. 
This figure shows that the small differences between the halo CMFs of \citet{Despali16} and those from the light cones or boxes (both using the infall masses for subhalos) translate into negligible differences in the SHMR. We are therefore confident that in our mass regime (i.e., for $\log(M_h/M_*)\geq 12.5$), the same relation holds both for the simulation and for the theoretical halo mass function.

Starting from low halo masses, the SHMR in the various redshift bins monotonically increases (with slope $\beta$) as a function of halo mass and reaches a peak at $M_A$, after which the relation decreases monotonically (with slope $\gamma$) at higher halo masses. The redshift evolution of the SHMR shows that above the characteristic halo mass $M_A$, the SHMR increases with increasing redshift, while this trend is reversed at lower masses. This can be interpreted as an evolution of the SFE, that is, the fraction of baryons locked in stars, which is defined as $f_{*}=(M_{*}/M_h)f^{-1}_b$,  where $f_b$ is the cosmological baryon fraction (i.e., $0.153$ for our Planck cosmology). When we assume a constant value of $f_b$ with redshift, then the value  of $(M_{*}/M_h)$ corresponds to the SFE. Fig.~\ref{SHMR_obsfit} therefore implies that the SFE is always lower than $35$\%  and increases for increasing redshifts at masses higher than the peak of the SHMR. This trend is reversed for masses below this limit. This means that massive galaxies (i.e., for halo masses higher than the SHMR peak) formed with a higher efficiency at higher redshifts (i.e., the downsizing effect; the term was coined by \citealt{Cowie96} to describe this behavior). In contrast, low-mass galaxies (i.e., for halo masses lower than the SHMR peak) formed with higher efficiency at lower redshifts.

It has been argued by several authors  \citep{Giodini09, Andreon10, Gonzalez13, Eckert16} that the baryon fraction is not constant with (halo) mass. The studies have focused mostly on local ($z\sim 0$) galaxy groups and clusters (i.e., for total masses of $\log(M_{\mathrm{500}}/M_\odot)\gtrsim 13$), where the total content of matter inside $R_{500}$ can be measured through X-ray studies. The general trend reported in these studies is that the baryon fraction increases with increasing mass and approximately reaches the universal baryon fraction at the highest measured masses ($\log(M_{\mathrm{500}}/M_\odot)\sim 15$). As an example, \citet{Gonzalez13} found a decrease in baryon fraction by a factor $\sim 1.5$ from $\log(M_{\mathrm{500}}/M_\odot)\sim 15$ to $\sim 14$.  
When we qualitatively assume that the baryon fraction decreases with decreasing halo mass throughout the entire range of halo masses we considered (from $\log(M_{h}/M_\odot)\sim 10.5$ to $15$), the SFE trend may be modified. In particular, if this scenario is in place, the SFE will still increase at masses higher than the peak of the SHMR, while below the peak, the decreasing trend we find may be less steep.

These studies are difficult to conduct, therefore the observed baryon fraction estimates are not available as a function of both redshift and halo mass. In order to explore whether the baryon fraction evolves with redshift, we can rely on hydrodynamic simulations, such as \citet{Dave09} and \citet{Dave10}. These authors ran set of simulations with a modified version \citep{Oppenheimer08} of the N-body+hydrodynamic code GADGET-2 \citep{Springel05}. Several physical processes were considered, such as star formation, radiative cooling, metal-line cooling, supernovae, kinetic outflows, and stellar winds. Dav\'e and collaborators found an approximately constant baryon fraction with halo mass at high redshifts (i.e., $z\sim 3$). At low redshift ($z\sim 0$), they found the same trend in mass as was reported in the studies mentioned above, that is, the baryon fraction increases with increasing halo masses and approximately reaches the universal baryon fraction at the highest halo masses ($\log(M_h/M_\odot)\sim 15$). The main result the authors find is that the baryon fraction decreases with increasing cosmic time at fixed halo masses (as an example, at $\log(M_h/M_\odot)=12,$ they find that the baryon fraction decreases by a factor $\sim 1.4$ from $z=3$ to $z=0$). This is particularly true at low halo masses (i.e., $\log(M_h/M_\odot)\lesssim 12$). If the baryon fraction were to also depends on redshift, this would lead to a modification of the values of SFE we find and would enhance the evolution from high to low redshifts at masses below the SHMR peak. 

To summarize, when we consider a constant baryon fraction with cosmic time and halo mass, we find that with increasing redshift, the SFE increases at masses higher than the peak of the  SHMR. This trend is reversed for masses below this limit. When we instead consider an evolution of the baryon fraction  with halo mass, cosmic time, or a combination of both, as predicted by some hydrodynamic simulations, the trends we find (with cosmic time and halo masses) are not affected, even though the precise values of the SFE might change. We also note that the same trends are preserved when we consider the model with {\it \textup{relative}} scatter.

\begin{table*}\label{tablefit}
\caption{ Best-fit parameters of the SHMR  for the reference case in Eq.~\ref{mosterrel} and their $68$\% confidence interval.}             
\label{indfit}      
\centering 
\begin{tabular}{ccccccccc} 
\hline\hline
$\Delta z$ & $A$ & $M_A$ & $\beta$ & $\gamma$ \\        \hline

$0.00\leq z < 0.20$&$0.0465^{+0.0015}_{-0.0015}$&$11.77^{+0.03}_{-0.03}$&$1.00^{+0.05}_{-0.05}$&$0.702^{+0.006}_{-0.006}$\\
$0.20\leq z < 0.50$&$0.0431^{+0.0025}_{-0.0025}$&$11.86^{+0.08}_{-0.07}$&$0.97^{+0.11}_{-0.09}$&$0.644^{+0.020}_{-0.019}$\\
$0.50\leq z < 0.80$&$0.0353^{+0.0015}_{-0.0014}$&$12.05^{+0.07}_{-0.07}$&$0.88^{+0.11}_{-0.10}$&$0.599^{+0.021}_{-0.019}$\\
$0.80\leq z < 1.10$&$0.0429^{+0.0018}_{-0.0017}$&$12.03^{+0.06}_{-0.05}$&$0.99^{+0.15}_{-0.13}$&$0.638^{+0.014}_{-0.014}$\\
$1.10\leq z < 1.50$&$0.0328^{+0.0013}_{-0.0013}$&$12.10^{+0.06}_{-0.06}$&$0.89^{+0.15}_{-0.13}$&$0.638^{+0.018}_{-0.016}$\\
$1.50\leq z < 2.00$&$0.0287^{+0.0008}_{-0.0007}$&$12.20^{+0.06}_{-0.05}$&$0.93^{+0.16}_{-0.14}$&$0.604^{+0.018}_{-0.017}$\\
$2.00\leq z < 2.50$&$0.0297^{+0.0006}_{-0.0006}$&$12.21^{+0.03}_{-0.03}$&$1.36^{+0.14}_{-0.13}$&$0.571^{+0.013}_{-0.012}$\\
$2.50\leq z < 3.00$&$0.0294^{+0.0010}_{-0.0009}$&$12.31^{+0.07}_{-0.06}$&$1.18^{+0.22}_{-0.19}$&$0.551^{+0.028}_{-0.025}$\\
$3.00\leq z < 4.00$&$0.0335^{+0.0021}_{-0.0020}$&$12.55^{+0.12}_{-0.10}$&$1.05^{+0.22}_{-0.18}$&$0.605^{+0.063}_{-0.052}$\\
\hline
\end{tabular}
\end{table*}

\begin{table*}\label{tablefit2}
\caption{ Best-fit parameters of the SHMR with relative scatter ($\sigma_R=0.2$ dex) in Eq.~\ref{mosterrel} and their $68$\% confidence interval.}             
\label{indfit_scat}      
\centering 
\begin{tabular}{ccccccccc} 
\hline\hline
$\Delta z$ & $A$ & $M_A$ & $\beta$ & $\gamma$ \\        \hline

$0.00\leq z <   0.20$&$  0.0494^{+  0.0018}_{-  0.0019}$&$  11.81^{+  0.03}_{-  0.03}$&$  0.94^{+  0.04}_{-  0.04}$&$  0.726^{+  0.006}_{-  0.006}$\\
$0.20\leq z <   0.50$&$  0.0429^{+  0.0026}_{-  0.0026}$&$  11.87^{+  0.06}_{-  0.06}$&$  0.99^{+  0.08}_{-  0.07}$&$  0.669^{+  0.016}_{-  0.015}$\\
$0.50\leq z <   0.80$&$  0.0348^{+  0.0016}_{-  0.0015}$&$  12.07^{+  0.06}_{-  0.06}$&$  0.86^{+  0.09}_{-  0.08}$&$  0.622^{+  0.017}_{-  0.015}$\\
$0.80\leq z <   1.10$&$  0.0429^{+  0.0019}_{-  0.0018}$&$  12.03^{+  0.04}_{-  0.04}$&$  1.04^{+  0.11}_{-  0.09}$&$  0.657^{+  0.011}_{-  0.011}$\\
$1.10\leq z <   1.50$&$  0.0325^{+  0.0013}_{-  0.0013}$&$  12.11^{+  0.05}_{-  0.05}$&$  0.87^{+  0.13}_{-  0.11}$&$  0.659^{+  0.014}_{-  0.013}$\\
$1.50\leq z <   2.00$&$  0.0285^{+  0.0008}_{-  0.0007}$&$  12.21^{+  0.04}_{-  0.04}$&$  0.94^{+  0.12}_{-  0.10}$&$  0.624^{+  0.014}_{-  0.013}$\\
$2.00\leq z <   2.50$&$  0.0297^{+  0.0006}_{-  0.0006}$&$  12.23^{+  0.03}_{-  0.02}$&$  1.31^{+  0.12}_{-  0.10}$&$  0.604^{+  0.010}_{-  0.009}$\\
$2.50\leq z <   3.00$&$  0.0294^{+  0.0009}_{-  0.0009}$&$  12.33^{+  0.06}_{-  0.05}$&$  1.13^{+  0.19}_{-  0.16}$&$  0.583^{+  0.023}_{-  0.020}$\\
$3.00\leq z <   4.00$&$  0.0330^{+  0.0018}_{-  0.0018}$&$  12.55^{+  0.10}_{-  0.09}$&$  1.05^{+  0.21}_{-  0.17}$&$  0.626^{+  0.045}_{-  0.038}$\\

\hline
\end{tabular}
\end{table*}

\subsection{Empirical best-fit model} 
\label{model}

\begin{figure}
   \centering
   \includegraphics[width=0.49\textwidth]{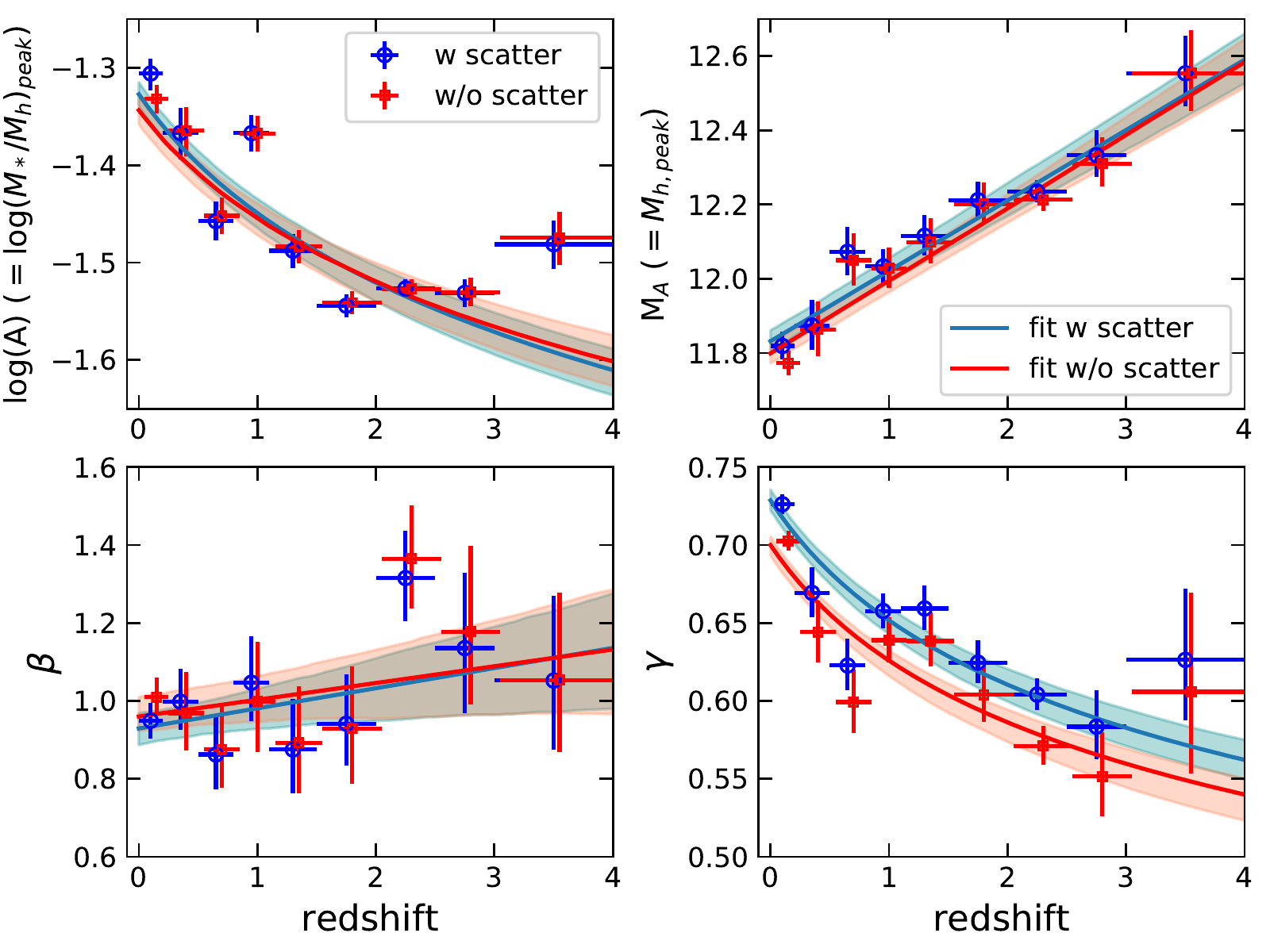}
  \caption{ Evolution of the parameters $M_A$, $A$, $\gamma,$ and $\beta$ (shown with points) for the case with and without {\it \textup{relative}} scatter in blue and red, respectively. We also show the empirical best-fit model and the $1\sigma$ uncertainties (shown with lines and shaded areas, respectively) in blue and red for the case with and without scatter, respectively. Red points have been shifted by $\Delta z=+0.05$ to facilitate visual comparison.}
              \label{evolparam}%
\end{figure}

In Fig.~\ref{evolparam} we show the evolution of the parameters of the fit at different redshifts for the cases with and without {\it \textup{relative}} scatter. In both cases, we note that below the characteristic halo mass $M_A$, the SHMR slope ($\beta$) is approximately constant with redshift. Conversely, for masses higher than $M_A$, the slope ($\gamma$) shows little evolution. As also explained in Sect.~\ref{obsSHMR}, the main difference between the results with or without the {\it \textup{relative}} scatter is in the value of $\gamma$. In particular, we find a systematic shift toward higher $\gamma$ values for the case that includes scatter.
This is a signature of the convolution we applied to introduce the {\it \textup{relative}} scatter to the SMF. 
The mass where the SHMR peaks ($M_A=M_{h,{\rm peak}}$) increases with redshift, while its normalization $A=(M_*/M_h)_{{\rm peak}}$ decreases with redshift.
After determining the nine best-fit parameter values in the different redshift intervals (listed in Table~\ref{indfit} and \ref{indfit_scat} for the two cases), we performed a fit on the evolution of the parameters $M_A$, $A$, $\gamma,$ and $\beta$ using Eqs.~\ref{1}, \ref{2}, \ref{3}, and \ref{4}. We show in Fig.~\ref{evolparam} the best-fit evolution of the parameters, along with its $1 \sigma$ uncertainties for the cases with and without  scatter.

\begin{figure}
   \centering
    \includegraphics[width=9cm]{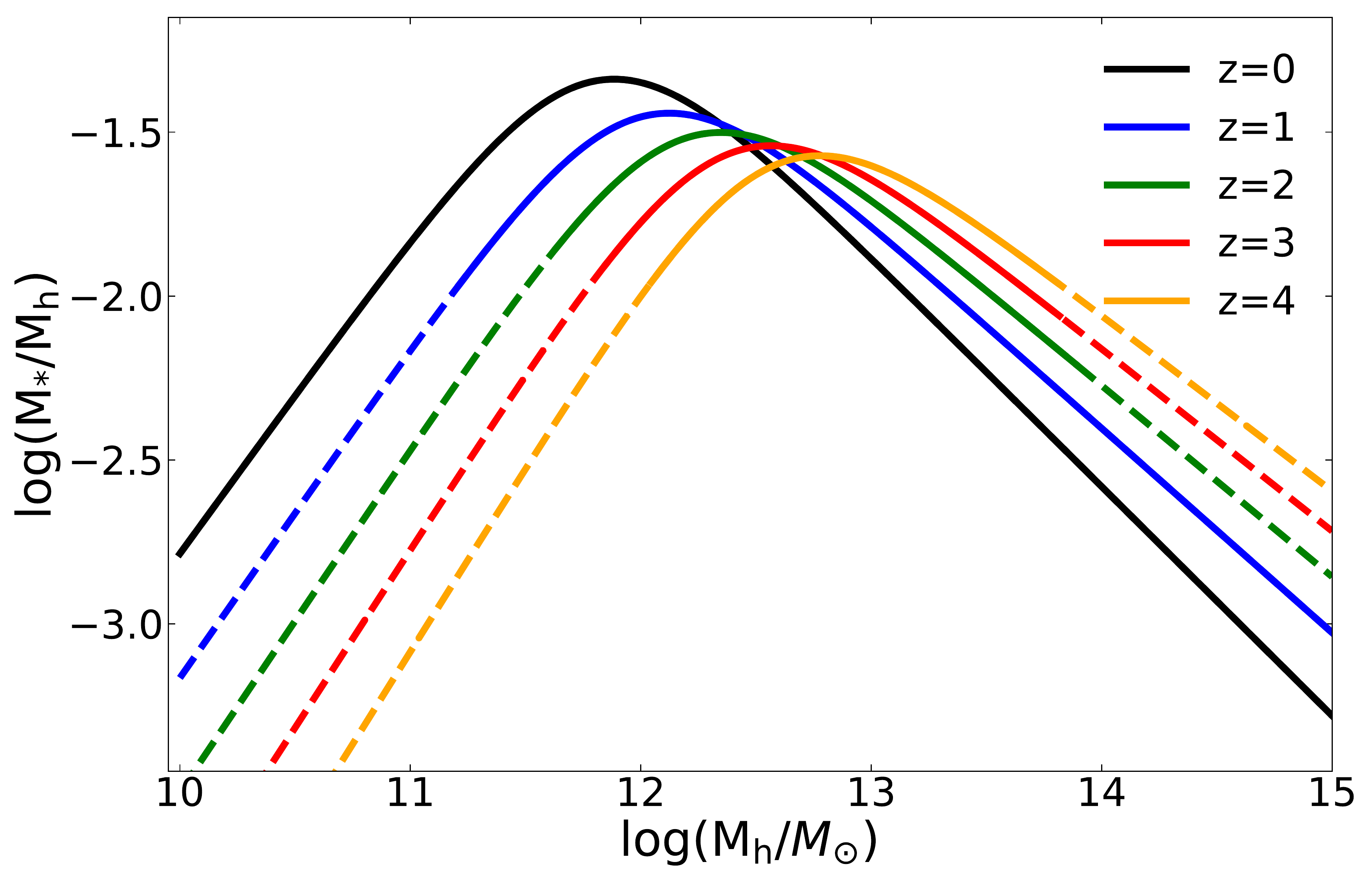}

    \includegraphics[width=9cm]{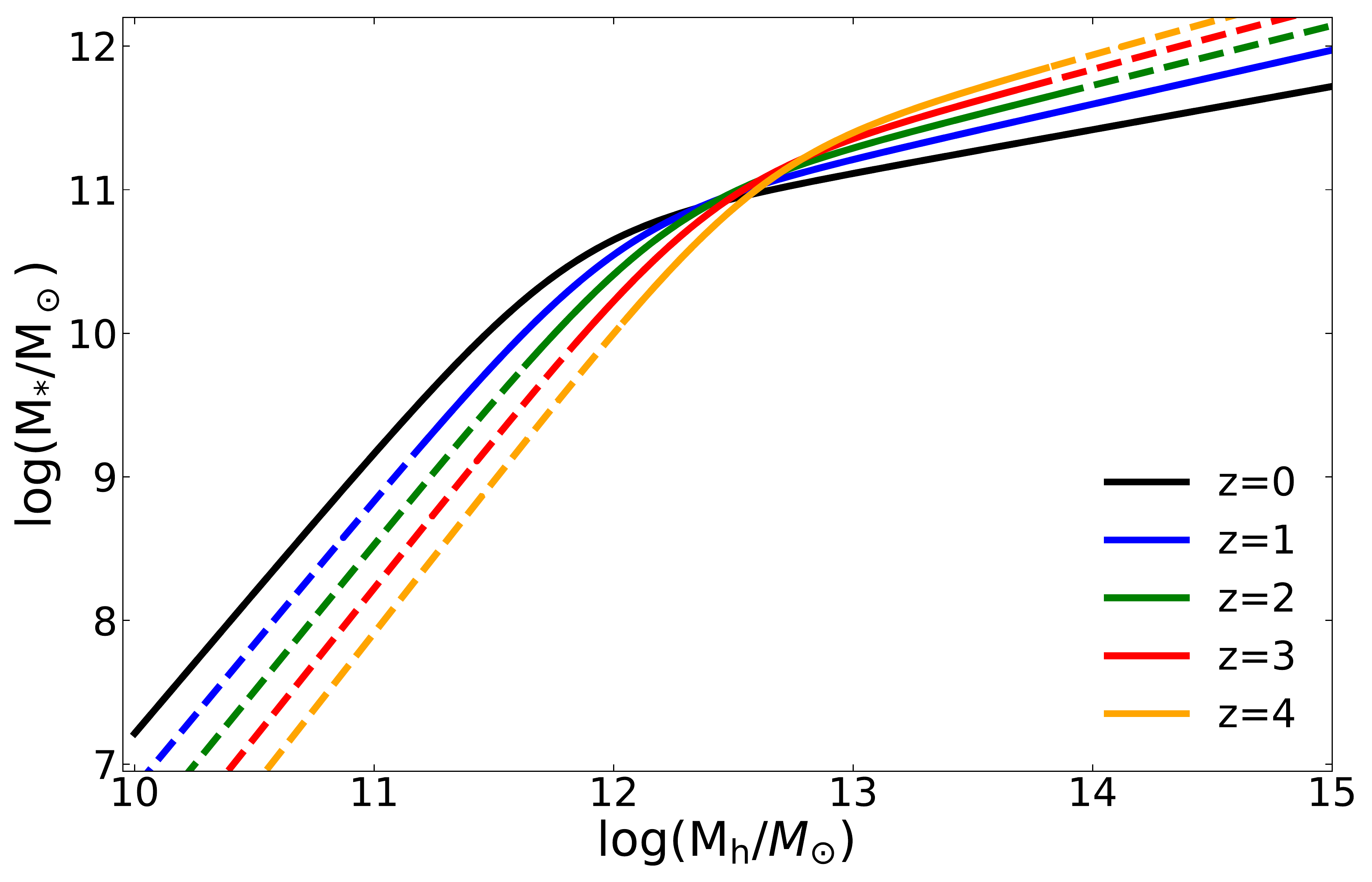}
  \caption{ Best-fit model plotted in a $\log(M_*/M_h)\,-\,\log(M_h)$ plane (top panel) and in a $\log(M_*)\,-\,\log(M_h)$ plane (top panel) for the reference case without scatter at different redshifts. Solid lines indicate the stellar mass range in the observed SMF,
  and the dashed lines represent the extrapolation of the model.}
              \label{SHMR_fit}%
\end{figure}

\begin{table*}
\caption{ Best-fit parameters of the SHMR evolution  for the reference case in Eqs.~\ref{1}, \ref{2}, \ref{3}, and \ref{4}, and their $68$\% confidence interval.}             
\label{table:1}      
\centering 
\begin{tabular}{ccccccccc}     
\hline\hline
 & $B$ & $\mu$ & $C$ & $\nu$ & $D$ & $\eta$ & $F$ & $E$ \\   \hline
Best fit&11.79&0.20&0.046&-0.38&0.709&-0.18&0.043&0.96\\
$1\sigma^+$&0.03&0.02&0.001&0.03&0.007&0.02&0.039&0.04\\
$1\sigma^{-}$&0.03&0.02&0.001&0.03&0.007&0.02&0.041&0.05\\
\hline
\end{tabular}
\end{table*}

\begin{table*}
\caption{ Best-fit parameters of the SHMR evolution with  {\it \textup{relative}} scatter($\sigma_R=0.2$ dex) in Eqs.~\ref{1}, \ref{2}, \ref{3}, and \ref{4}, and their $68$\% confidence interval.}             
\label{table:2}      
\centering 
\begin{tabular}{ccccccccc}     
\hline\hline
 & $B$ & $\mu$ & $C$ & $\nu$ & $D$ & $\eta$ & $F$ & $E$ \\   \hline
Best fit&11.83&0.18&0.047&-0.40&0.728&-0.16&0.052&0.92\\
$1\sigma^+$&0.03&0.02&0.001&0.03&0.007&0.01&0.034&0.04\\
$1\sigma^{-}$&0.03&0.02&0.001&0.03&0.007&0.01&0.036&0.04\\

\hline
\end{tabular}
\end{table*}

Figure~\ref{SHMR_fit} shows our derived best-fit models for our reference case without the {\it \textup{relative}} scatter, which links the halo mass to the ratio between stellar mass and halo mass model as a function of halo mass at various redshifts. In Tables~\ref{table:1} and \ref{table:2} we report the best fit and the $68$\% confidence interval for the eight parameters of the fit for the case with and without scatter, respectively.
Fig.~\ref{evolparam} shows that the models predict an evolution of the parameters with redshift that smoothes out all features characteristic of the COSMOS field, such as the well-known overdensity located at $z \sim 0.7$ (e.g., \citealt{McCracken15}). This overdensity  shifts the high-mass end of the SMF to higher stellar masses and therefore shifts the parameter $M_A$ of the best fit to higher values (and consequently, $\gamma$ and $\beta$ to lower values). 
We note that for the last redshift bin (i.e., $3.0\leq z<4.0$), the values of the normalization ($A$) and the slope at high masses ($\gamma$) show large error bars (compared to the other redshift bins) and are not well represented by the best-fit models. However, as we describe below, both models are still able to reproduce (with large error bars) the observed SMF trends.

\subsection{Comparison with other works}
\label{literature}

\begin{figure*}[h]
   \centering
   \includegraphics[width=14cm]{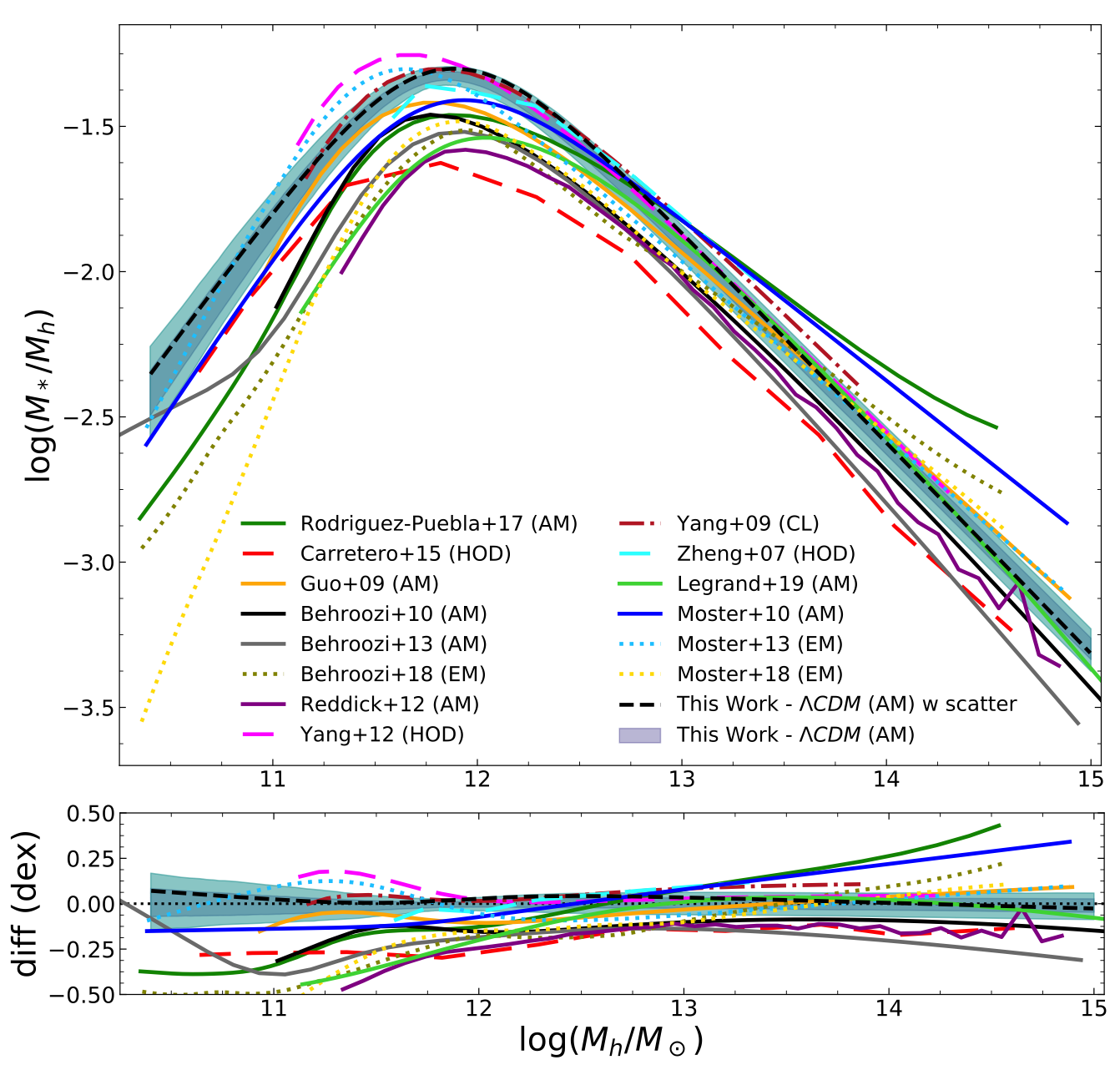}
  \caption{Best-fit SHMR compared to previous results at $z\sim 0$. { Dark and light shaded areas represent the $1$ and $2\sigma$ errors, respectively, for our reference case without scatter at redshift $0<z<0.2$.  For the case with scatter, we show our results with a dashed black line without errors (which are comparable to errors of the case without scatter).} The comparison includes results from abundance matching (AM; \citealt{Behroozi10,Behroozi13,Moster10,Rodriguez17}, empirical modeling (EM; \citet{Moster10,Moster18,Behroozi19}), halo occupation distributions (HOD; \citet{Carretero15,Yang12,Zheng07}) and clusters selected from SDSS spectroscopic data (CL;\citealt{Yang09}). In the bottom panel we show the logarithmic difference between our results and the other literature works. The shaded area represents our errors.}
              \label{SHMR_HOD_z0}%
\end{figure*}

\begin{figure*}[h]
{\includegraphics[width=0.49\textwidth]{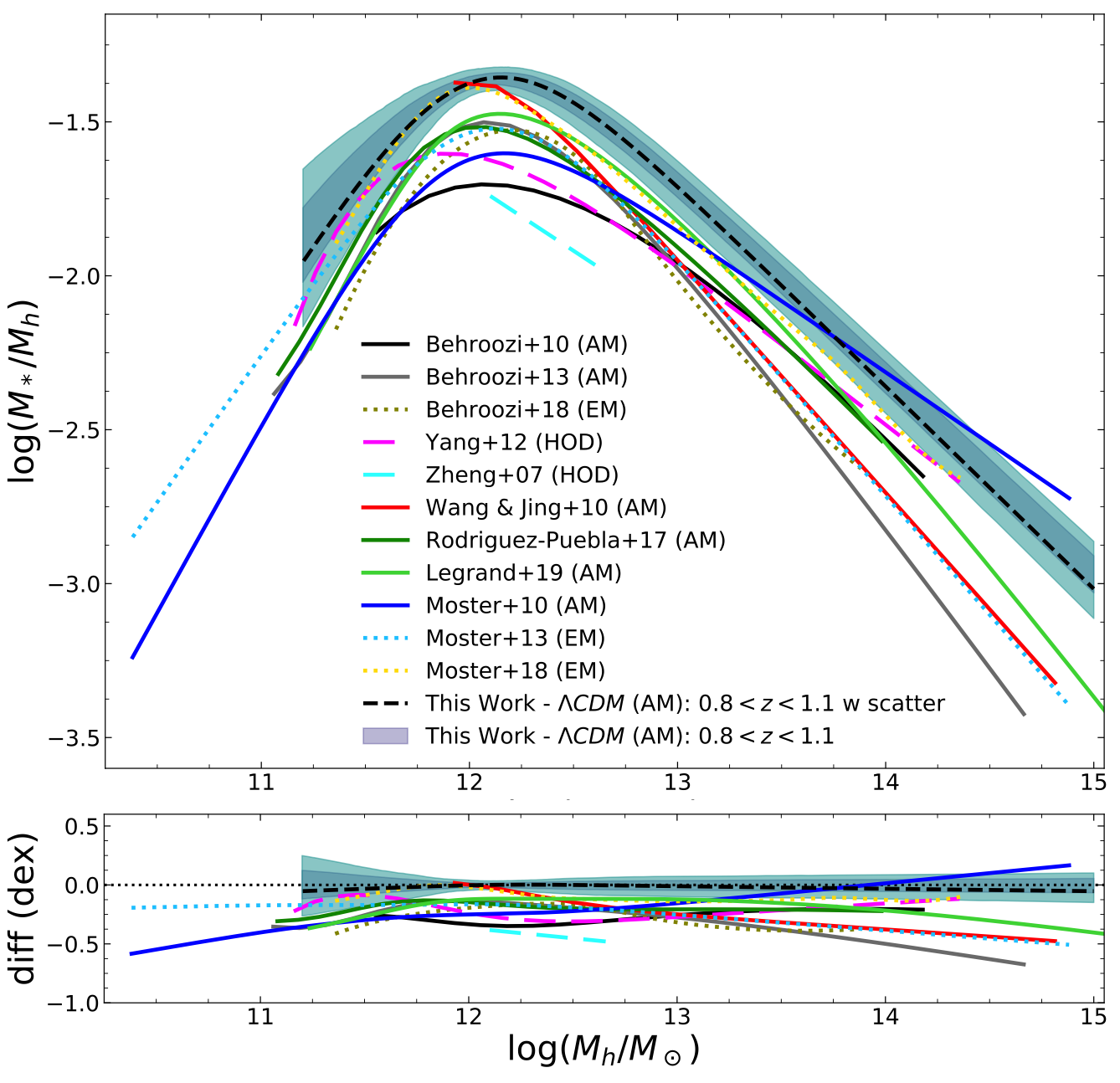}
\includegraphics[width=0.495\textwidth]{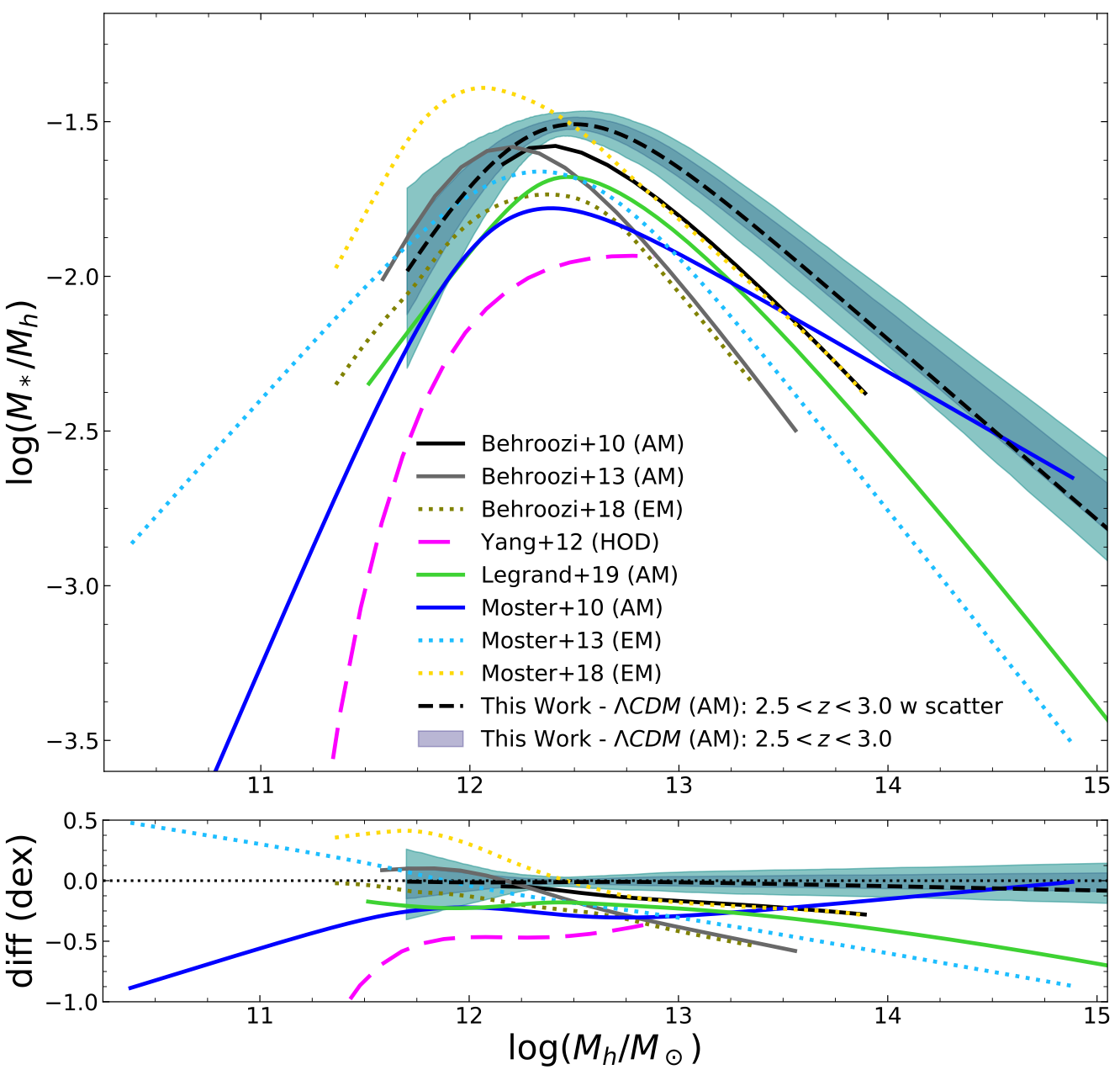}}
\caption{Best-fit SHMR compared to previous results at $z\sim 1$(left) and $z\sim 3$ (right). Symbols are the same as in Fig.~\ref{SHMR_HOD_z0}.}
\label{SHMR_HOD_z13}
\end{figure*}

In Fig.~\ref{SHMR_HOD_z0} we show a comparison of our best-fit SHMR (with and without scatter) with several literature results at $z \sim 0$, and a comparison at $z\sim 1$ and $z\sim 3$ is shown in Fig.~\ref{SHMR_HOD_z13}. For the {\it \textup{reference}} case without scatter, our results are shown along with the corresponding $1$ and $2\sigma$ errors of the fit in the redshift bin indicated in the plots (i.e., $0.0<z<0.2$ for $z\sim 0$, $0.8<z<1.1$ for $z\sim 1$ and $2.5<z<3.0$ for $z\sim 3$).

Performing a comparison with other works is not always straightforward because other papers have often made different assumptions on the cosmological model, on the definition of halo mass, or on the measurement of stellar mass\footnote{Most of the other literature data were kindly provided by Peter Behroozi and are also shown in several other papers (e.g., \citealt{Behroozi10,Behroozi13,Behroozi19}). Corrections for differences in the underlying cosmology have been applied by \citet{Behroozi10} using the process detailed in their Appendix A. We also made use of the semiautomated tool \textit{WebPlotDigitalizer} to extract datapoints from literature plots.}.
The assumptions used to derive stellar masses have not been adjusted instead because such adjustments can be complex and difficult to apply using simple conversions.
We only converted the IMF of all stellar masses into that of the \citet{Chabrier03} and \citet{B&C03} stellar population synthesis model.
Additionally, we converted all quoted halo masses into $M_{200}$ masses as defined in Sect.~\ref{halocat} by assuming a \citet{Navarro97} profile (NFW) and by calculating the correction between our halo mass definition and the mass definitions used in other works.
For the sake of clarity, we do not show errors of other literature works in the figures, but they can be found in \citet{Behroozi10}. 

In general, we find a large spread among all literature works that increases with increasing redshift. Most of the cited works show some differences  with respect to our results at all redshifts. At $z=0$ our results agree with literature  within the dispersion at intermediate halo masses, while differences can become larger than $0.3$\,dex (a factor $\sim 2$) at low and high $M_h$ ($\log(M_h/M_\odot)<11.5$ and $\log(M_h/M_\odot)>14$).  The works with the smaller differences compared to our results are \citet{Yang09} and \citet{Yang12}, and the \citet{Carretero15} model has the largest differences. At higher redshifts the dispersion between different works increases and our results are in general higher than other literature works, with differences almost always, but with some exceptions, higher than $0.3$\,dex in most of the halo mass range. At $z=3$ most of the works do not agree with each other, even up to $\sim 1$\,dex. In Appendix~\ref{appendix1} we provide some details on the several works with which we compared our results. 

It is evident that there are still discrepancies between different works in literature, and this is particularly true at higher redshifts. 
Part of these differences are due to the different methods that were used. The remaining differences can be ascribed to the SMFs (or luminosity functions) that were adopted to derive the relation, or to different halo finders. As shown in \citet{Knebe11}, different halo finder algorithms may obtain halo masses that are different by up to $10\%$ at $z\sim 0$, and this difference increases with increasing redshift. We adopted the estimates derived from COSMOS for the SMF, which is the largest field observed so far with a continuous and homogeneous coverage in redshift from $z=0.2$ to $z=4$. This field also has one of the best statistical and photometric accuracies in photometric redshifts, stellar masses, and SMFs determination for such a wide redshift range. In addition, we note that our result including {\it \textup{relative}} scatter in stellar mass at fixed halo mass lies at all redshifts in the $2\sigma$ confidence region of the case without scatter.

\subsection{Comparison with semianalytic models}\label{semi-anal}

\begin{figure}[ht]
   \centering
   \includegraphics[width=9cm]{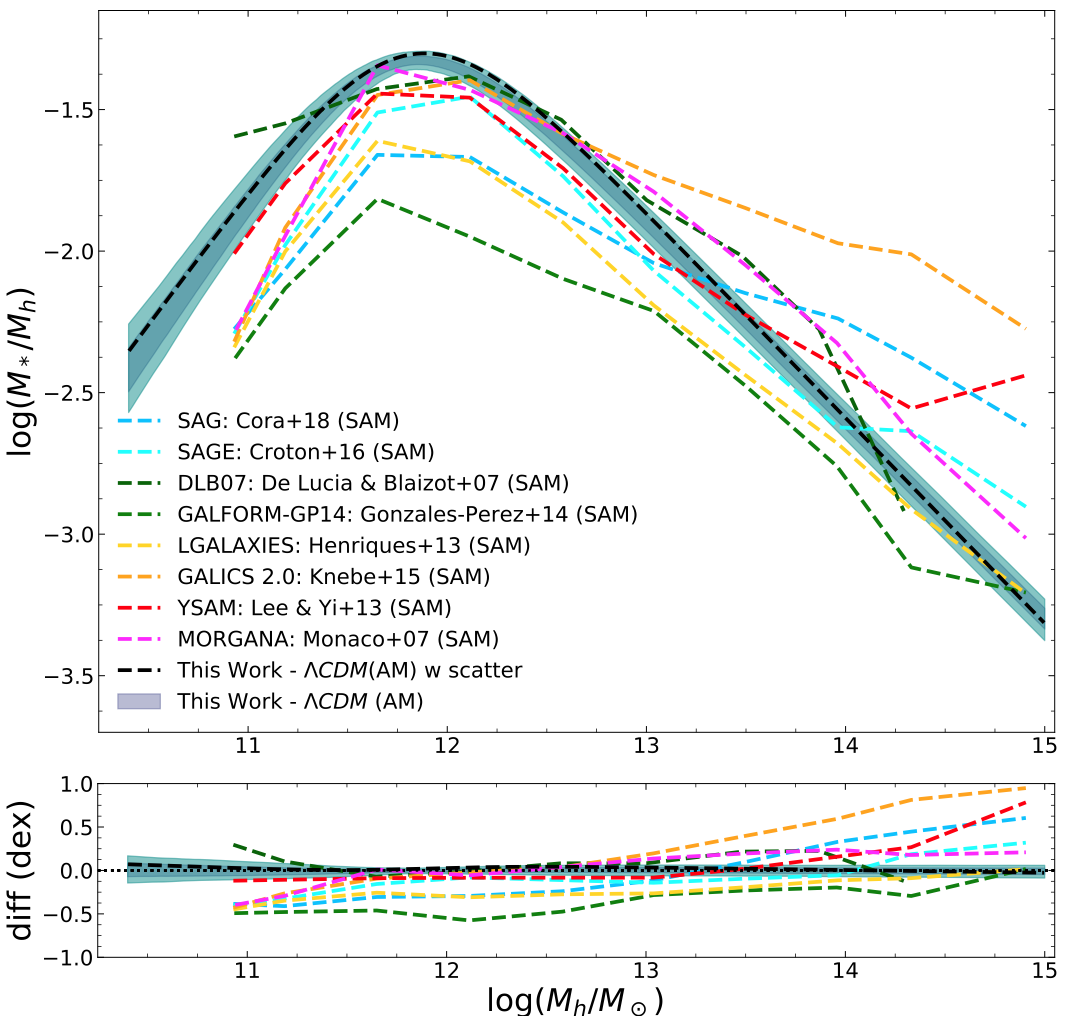}
  \caption{Best-fit of the SHMR compared to semianalytic models at $z\sim 0$ from the COSMIC CARNage project \citep{Knebe18}. Dark and light shaded areas represents the $1$ and $2\sigma$ errors, respectively, of the case without scatter. The black dashed line represents our result with scatter.}
              \label{SHMR_SAMs_z0}%
\end{figure}

\begin{figure*}
   \centering
{\includegraphics[width=0.49\textwidth]{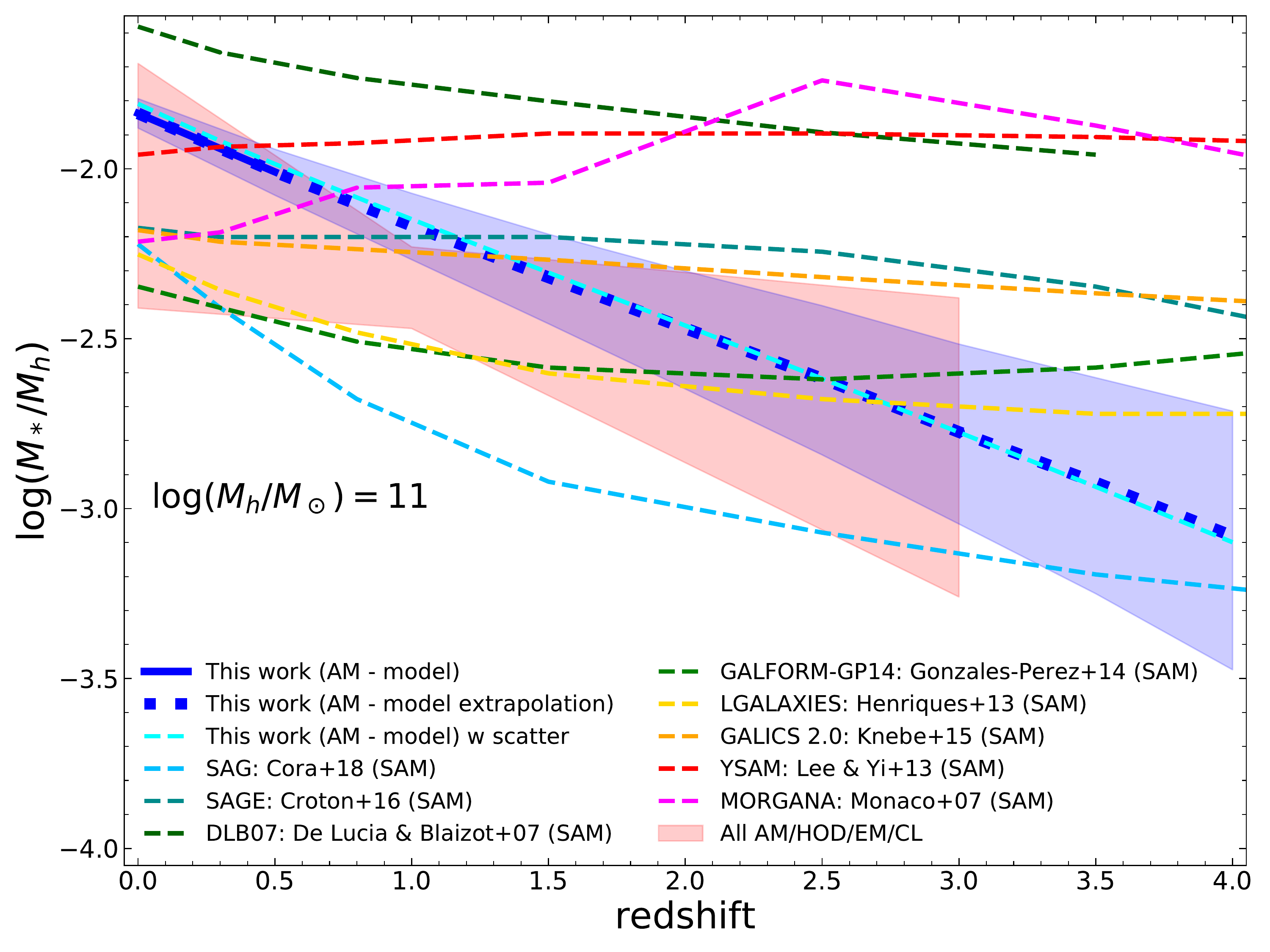}
\includegraphics[width=0.49\textwidth]{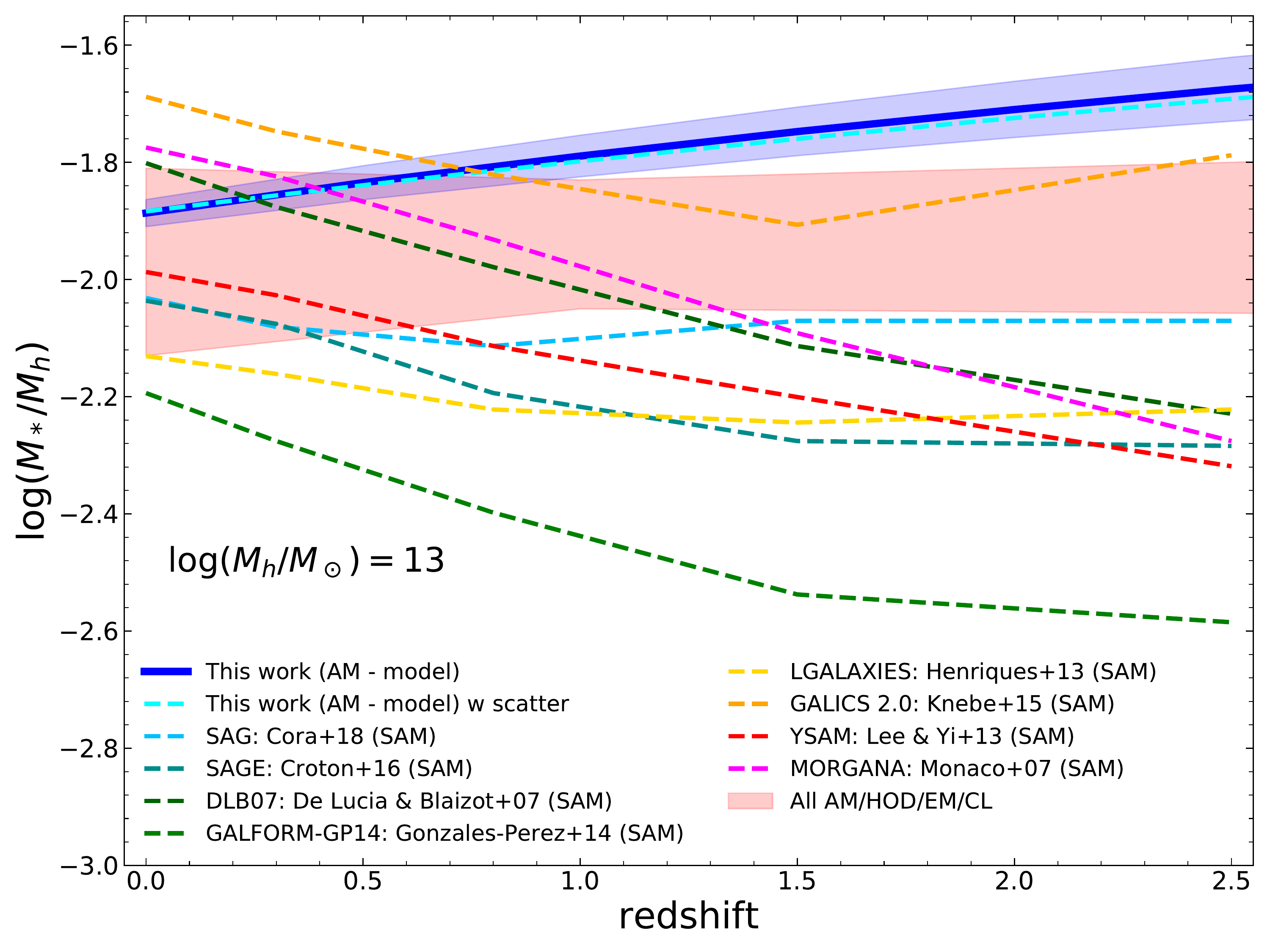}}
  \caption{Evolution of the SHMR at fixed halo masses of $\log(M_h/M_\odot)=11.0$ (\textbf{left}) and $\log(M_h/M_\odot)=13.0$ (\textbf{right}). Our results derived from the model described in Sect.~\ref{model} are shown with a blue line, and the corresponding shaded area represents $1\sigma$ uncertainties for our {reference} case. Cyan dashed lines show our results with {\it \textup{relative}} scatter (errors are not shown, but are comparable to the errors of the {reference} case). Solid lines indicate the observed mass range in the SMF, and the dotted line represents the range of masses extrapolated by the model. 
  Dashed colored lines represent the eight SAMs of the Cosmic CARNage project. The evolution in the left panel is shown from $z=0$ to $z=4$, and the right panel only shows the evolution up to $z=2.5$ because data at higher redshifts for these halo masses are lacking in the Cosmic CARNage simulations. Red shaded regions show the range covered by all the AM, HOD, EM, and CL results, as presented in Figs.~\ref{SHMR_HOD_z0} and \ref{SHMR_HOD_z13} and described in Sect.~\ref{literature} and Appendix~\ref{appendix1} (without the results of this work). We note that not all the works in literature extend their results down to $\log(M_h/M_\odot)=11$ or up to $\log(M_h/M_\odot)=13,$ and we therefore show only those whose results reach these masses.}
\label{SHMR_evol_SAMs}%
\end{figure*}

In Fig.~\ref{SHMR_SAMs_z0} we compare our best-fit result at $z\sim 0$ to the results of eight SAMs of galaxy formation and evolution. These SAMs were run on the same underlying CDM simulation (cosmological box of comoving width $125\,\mathrm{h}^{-1}$ Mpc) and the same merger trees within the Cosmic CARNage project \citep{Knebe18, Asquith18}. The SAMs are \textit{SAG} by \citet{Cora18,Cora19}, \textit{SAGE} by \citet{Croton16}, \textit{DLB07} by \citet{Delucia07}, \textit{GALFORM-GP14} by \citet{Gonzalez14}, \textit{LGALAXIES} by \citet{Henriques13,Henriques15,Henriques17}, \textit{GALICS 2.0} by \citet{Knebe15}, \textit{YSAM} by \citet{LeeYi13}, and \textit{MORGANA} by \citet{Monaco07}. The galaxy formation models used in the project are described in \citet{Knebe15,Knebe18}, where all of them are summarized in a concise and unified manner, along with their main features and differences. However, it is worth mentioning that \textit{\textup{all}} models have been calibrated with SMFs at $z=0$ \citep{Baldry08,Li&white09,Baldry12}, SMFs at $z=2$ \citep{Dominguez11, Muzzin13, Ilbert13, Tomczak14}, star formation rate function at $z=0.15$ by \citet{Gruppioni15}, cold gas mass fraction at $z=0$ evaluated by \citet{Boselli14}, and the relation of black hole to bulge mass \citep{McConnell13,Kormendy13}. In order to further align the various galaxy formation models, they have all assumed a Chabrier IMF, a metallicity yield of $0.02,$ and a recycled fraction of $0.43$. Moreover, a \textit{Planck} cosmology was used, and the halo mass was defined as $M_{200}$, as in this work. Results from the Cosmic CARNage simulations therefore do not require any rescaling of the properties in order to allow a comparison with our results.
All the considered semianalytic galaxy formation models populate the dark matter halos with galaxies whose properties depend on the details of the formation history of the halo in which they are placed. Subsequent galaxy evolution then shapes the galaxy SMF and therefore the SHMR.  
Fig.~\ref{SHMR_SAMs_z0} shows that the scatter at $z\sim0$  is already large between the SHMR of different SAMs, and this is particularly evident in the position of the peak of the relation. We investigate this in more detail in the following section. 
From the comparison, we find that none of the eight models reproduces the observed SHMR we propose at $z=0$ at any masses. The large differences between our result and the SAMs (which at maximum is up to a factor of $\sim 8$ in $(M_*/M_h)$ at $\log(M_h/M_\odot)=15.0$ when our result is compared to \textit{GALFORM-GP14}) reflect the intrinsic difficulty of treating the physical processes related to galaxy formation and evolution. 

\subsection {Evolution of the SHMR}

In Fig.~\ref{SHMR_evol_SAMs} we show the evolution of the SHMR as a function of redshift at two fixed halo masses for both cases (with and without scatter): one located at masses below the peak ($\log(M_h/M_\odot)=11.0$), and the other above it ($\log(M_h/M_\odot)=13.0$). Figs.~\ref{evolparam} and \ref{SHMR_fit} have clearly shown that the redshift evolution has completely opposite trends at these two masses,
similarly to other  AM, HOD, EM, and CL literature results. We have described before that results with or without scatter are very similar.
At $\log(M_h/M_\odot)=11.0$, our SHMR shows a strong evolution (more than one order of magnitude) of its value (even \textbf{if} large errors are associated to the parameter $\beta$), with a decreasing trend with increasing redshift. 
Assuming a simple hierarchical structure formation scenario, we would expect that $M_*/M_{h}$ remains constant with redshift. However, baryons do not share the bottom-up evolution of dark matter halos, as proven by several results (e.g., \citealt{Cowie96,Fontanot09,Thomas10}). The most massive galaxies (mainly early-type galaxies hosted in galaxy groups and clusters) are dominated by old stellar populations, and appear to have formed their stellar mass relatively quickly at the beginning of their life, while faint field galaxies (usually late-type galaxies) appear to have continued to actively form stars over the last billion years, and their stellar population is dominated by young stars. This is the so-called downsizing scenario. 

For SAMs, the value of the SHMR appears not to evolve as much as we find. Moreover, different SAMs disagree whether the SHMR value decreases, as in our work (e.g. \textit{SAG}, \textit{DLB07}, and \textit{LGALAXIES}), or remains approximately constant (e.g., \textit{YSAM} and \textit{GALFORM-GP14}). This indicates that the downsizing effect is not well reproduced by the evolution of $M_*/M_h$ in all analyzed SAMs. This might be due to the well-known overcooling problem (e.g., \citealt{Benson03}): galaxies are modeled to form as gas cools inside of dark matter halos \citep{White78}. However, the mass function of dark matter halos rises steeply at low masses \citep{Reed07}. Because cooling is very efficient in these low-mass halos, the galaxy mass and/or luminosity function are expected to show a similar slope at the low-mass or low-luminosity end. The observed slopes are instead much shallower (as a consequence of the downsizing effect), and therefore some form of feedback is postulated to mitigate this discrepancy, typically from supernovae and AGN. However, these feedback mechanisms are still not fully understood and their modeling is accordingly uncertain. We also note that the model \textit{MORGANA} is the only SAM that shows an increasing trend with increasing redshift. This may be due to the resolution of the simulation and/or an excessive overcooling that probably is due to the treatment of the feedback mechanisms. In a future work we will perform further comparisons with more recent SAMs that have improved feedback effects at low stellar mass to better reproduce SMF \citep{Hirchmann16,Delucia17,Zoldan19}, and hydrodynamic simulations \citep{Nelson15}, which are not included in  Cosmic CARNage.

At $\log(M_h/M_\odot)=13.0$, we find an increasing value of the SHMR with increasing redshift, in accordance with other similar AM, HOD, EM, and CL literature results, and its evolution becomes even stronger at higher halo masses (shown in Fig. \ref{SHMR_fit}). All SAMs instead show an opposite decreasing trend (with the exception of \textit{SAG} and \textit{LGALAXIES,} which show an approximately flat trend). We show their evolution only up to $z=2.5$ because SAMs lack data at higher redshifts at this halo mass as a consequence of the relatively small volume of the simulation, which does not contain such rare massive halos at $z>2.5$. 
These different trends may be an indication that some physical processes are not yet accurately modeled in SAMs, such as feedback mechanisms that may affect the build-up of stellar mass in galaxies. However, the box size of the underlying cosmological simulation for dark matter ($125\,\mathrm{Mpc}\,h^{-1}$ on a side) is too small to carry out a definitive comparison between SAMs and other results such as ours.

\begin{figure*}
   \centering
{\includegraphics[width=0.49\textwidth]{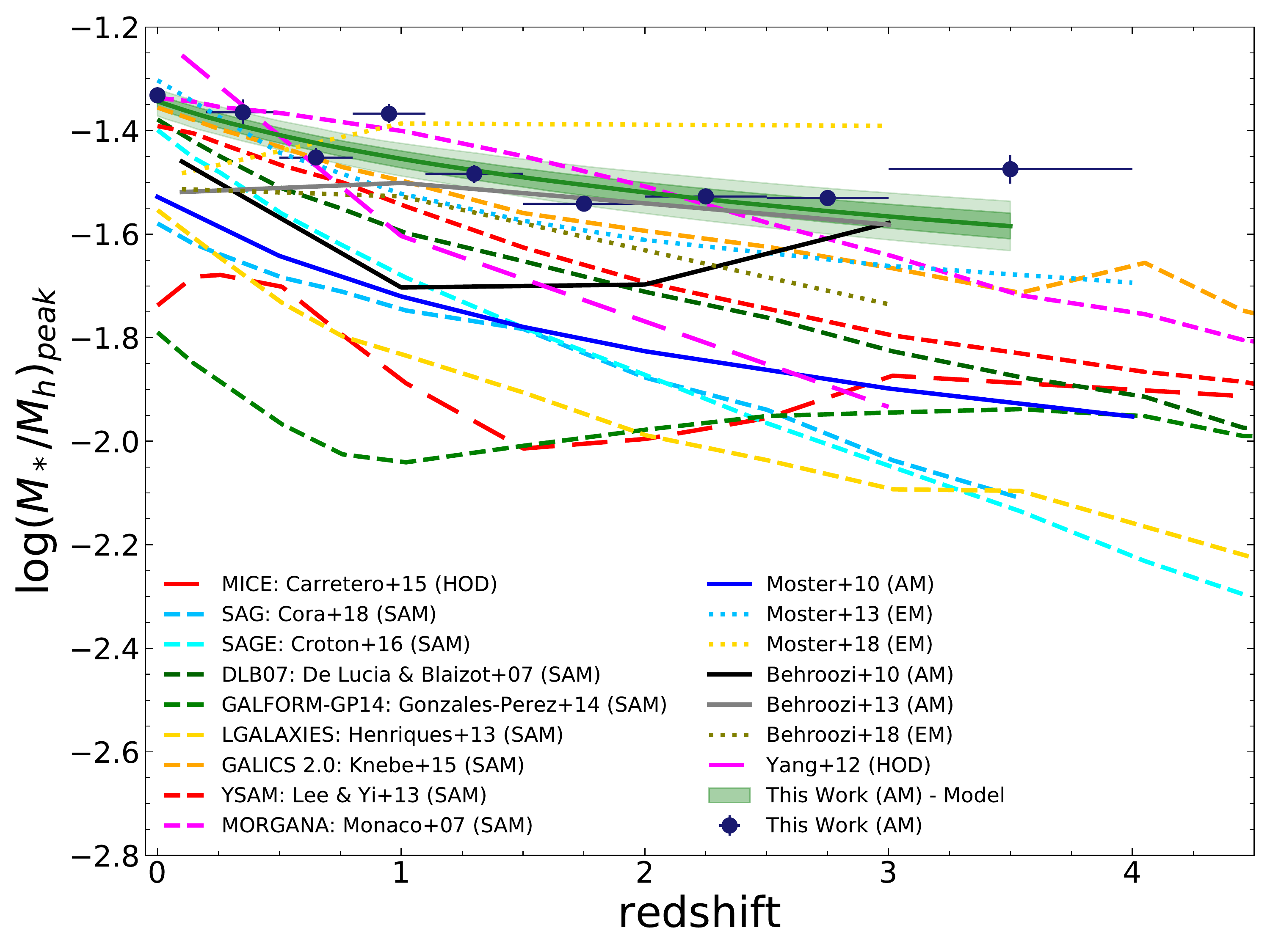}
\includegraphics[width=0.49\textwidth]{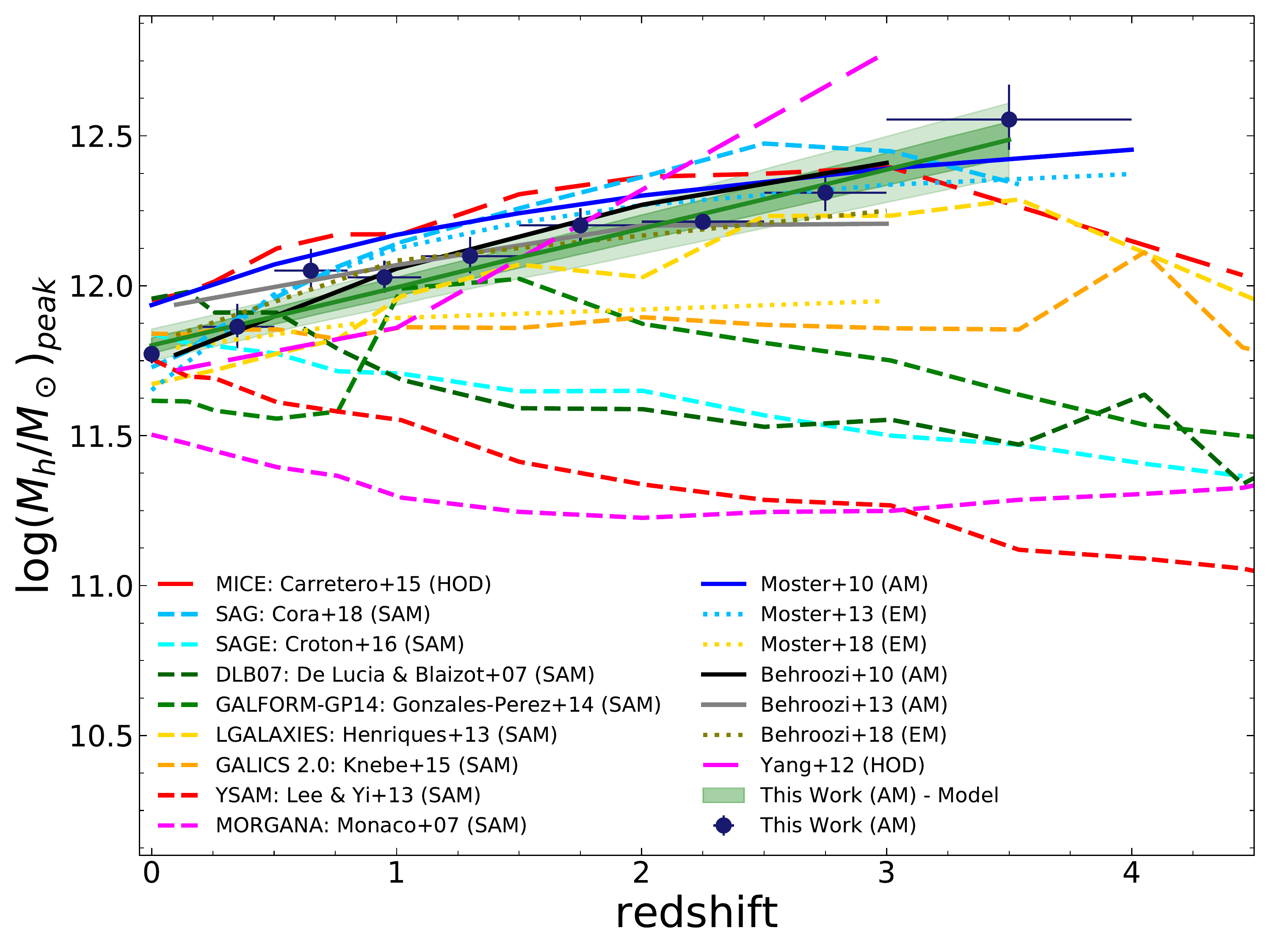}}
  \caption{\textbf{Left:} Evolution of the peak of the SHMR( for our reference case) compared with previous literature works and results of SAMs of galaxy formation and evolution. Dark blue points and error bars represent the fit of our measured SHMR.  The model we propose is shown in green, along with $1$ and $2\sigma$ errors estimated on the posterior distribution of the free parameters of the fit. \textbf{Right:} Comparison of the evolution of the halo mass at the peak of the SHMR( for our reference case). Symbols are the same as in the other panel. }
              \label{SHMR_PEAK}%
\end{figure*}
\subsection{Evolution of the SHMR peak}\label{peak}

In Fig.~\ref{SHMR_PEAK} we show the redshift evolution of the peak halo mass $M_{h,{\rm peak}}$ and of the corresponding $(M_*/M_{h})_{\rm peak}$ value. In this case, we only show results for our reference case without  {\it \textup{relative}} scatter because the differences of the peaks in the two cases are negligible (as also shown in Fig.~\ref{evolparam}). We estimate them from the best fit of the observed SHMR shown in Fig.~\ref{SHMR_obsfit}, whose uncertainties are mainly due to errors in the SMFs (i.e., photometric redshift errors, stellar mass errors, cosmic variance, and Poissonian errors).
The uncertainties on the SHMR  are evaluated from the 16th and 84th percentiles of the posterior distribution of the fitting procedure (see Fig.~\ref{SHMR_obsfit}).
Moreover, we also show the same quantities derived from our model, that is, $\log(M_*/M_h)_{\rm peak}(z)=\log(A$(z)) and $\log(M_h/M_\odot)_{\rm peak}(z)=\log(M_A$(z)). The evolution of $\log(M_h/M_\odot)_{\rm peak}$ is also well represented by the model, while the model is more different from $\log(M_*/M_h)_{\rm peak}$ in the highest redshift bin (i.e., $3.0\leq z<4.0$) and at intermediate and low redshift (i.e. $0.8\leq z<1.1$). We showed in Fig.~\ref{evolparam} and explained in Sect.~\ref{model} that the model tends to smooth out all peculiar features of the COSMOS field resulting in a smooth evolution of the parameters. 
We also show several results from the literature (described in Sect.~\ref{literature} and Appendix~\ref{appendix1}) and SAMs (presented in Sect.~\ref{semi-anal}). We do not show the errors of other works for clarity. The errors are comparable to the errors we show for our results.

Fig.~\ref{SHMR_PEAK} shows that we observe an approximately flat trend of the evolution of the $(M_*/M_{h})_{\rm peak}$ value, with only a slight decrease by a factor of $\sim 1.4$ from $z=0$ to $z=4$, although the uncertainties at high redshift are quite large.
At $0.8<z<1.1$, however, our data show an anomalous increase that is due to the SMFs we adopted. Fig.~\ref{CMFs} also showed that the CMFs show a shift toward higher stellar masses with respect to the general evolution of the other redshift bins. This may be due to an overdensity of massive galaxies in the COSMOS field or to errors in the SED-fitting procedure at these redshifts. In a future paper we will perform a parallel analysis using different SMFs (e.g., \citealt{Davidzon17}) and cosmological simulations (e.g., other simulations from the {\small DUSTGRAIN-}{\it pathfinder} set) in order to assess the reliability of the results we present. 

We find in general that even though the general trend of the evolution of the SHMR peak is different in the literature, most of the works find that the value of $(M_*/M_{h})_{\rm peak}$ decreases with increasing redshift (e.g., \citealt{Moster10,Moster13,Yang12}) and only few works find that it remains approximately constant (e.g., \citealt{Behroozi13,Moster18}). Moreover, there is still a large scatter between literature results (up to a factor of $\sim 2$ in $M_*/M_h$), which leaves this an open question. We observe a general trend with the $(M_*/M_{h})_{\rm peak}$ value decreasing with increasing redshift in differnt SAMs for all but \textit{GALFORM-GP14}, which presents a minimum a $z\sim0.9$ and remains flat thereafter.

As we discussed above, the value of $(M_*/M_{h})_{\rm peak}$ can also be interpreted as the peak of SFE.
The value of $(M_*/M_{h})_{\rm peak}$ indicates that galaxies with these characteristic masses are in general the most efficient at turning baryons (gas) into stars during their lifetime. We find that the peak of the SFE ranges from $\sim 0.35$ to $\sim 0.3 \, \pm 0.04$ (going from $z=0$ to $z=4$), which means that $\sim 30-35 \%$ of all available gas has been turned (and remain locked) into stars; the peak efficiency lies at $z\sim 0$. We recall here that a high fraction of gas that is processed in stars returns to the ISM at the end of stellar evolution (up to $30-50$\% depending on the age, star formation history, and IMF, e.g., up to $48$\% for a Chabrier IMF, \citealt{B&C03}). When we also account for the return fraction $R$ in calculating the SFE (including all the gas involved in the star formation), the values we find would therefore be $(1-R)^{-1}$ times higher (up to $\sim 1.9$ for a Chabrier IMF).

These low efficiencies suggest that most of the baryons are in the form of interstellar and intergalactic diffuse gas. A recent study by \citet{Posti19} found that the SFE for spiral galaxies   is a monotonically increasing function of $M_*$ and that massive spirals (i.e., $M_*\gtrsim 10^{11} M_\odot$) have $f_*\sim 0.3-1$. This indicates that these systems have turned all their available gas into stars. However, this result has been derived only on local massive spirals. 
Several works have suggested that the shape of the SHMR depends on the galaxy type (e.g., \citealt{Mandelbaum06, Conroy07,Rodriguez15}), with red and passive early-type galaxies residing in most massive halos with respect to blue late-type galaxies. However, it is not straightforward to precisely determine from a simulation which halos host late- or early-type galaxies because the halo masses of these two classes of objects overlap at approximately the peak of the SHMR. We postpone this analysis to a future work.

The evolution of $M_{h,{\rm peak}}$ contains useful information as well. The value of $M_{h,{\rm peak}}$ approximately coincides with the knee of the SMF, with an offset depending on the slopes of the HMF and SMF. Star formation is therefore most effective and least influenced by either stellar or AGN feedback for these halo masses \citep{Moster10,Yang12}. 
We here find an increasing value of $M_{h,{\rm peak}}$ with increasing redshift. The value of $M_{h,{\rm peak}}$ corresponds to the most efficient star formation. We can therefore infer that the trend we find may be a signature of the downsizing effect. In other words, we find that the halo mass for which star formation is most efficient monotonically decreases with cosmic time. Massive galaxies were therefore formed at earlier times than less massive objects, even if $(M_*/M_{h})_{\rm peak}$ slightly increases with cosmic time. We showed in Fig.~\ref{SHMR_PEAK} (right panel) that some literature works (e.g., \citealt{Yang12,Moster10,Moster13,Behroozi10,Behroozi13}) have reported the same trend with redshift, while others (e.g., \citealt{Carretero15}) and SAMs do not show the same trend. In general, SAMs show lower values; some of them (\textit{SAGE}, \textit{GALFORM}, \textit{YSAM}, and \textit{DLB07}) show a decreasing trend with increasing redshift, with the exception, at least marginally, of \textit{SAG} and \textit{LGALAXIES}, which yield a rising $M_{h,{\rm peak}}$ value up to redshift $z \sim 3$. However, the scatter between different studies is large in this case as well (up to a factor of $\sim 3$ in $M_h$, and it increases with redshift). This implies that much work remains to be done in order to precisely determine the SHMR at all redshifts and understand the physical processes that give rise to it.

\section{Testing the SHMR}\label{testingMF}

\begin{figure*}
   \centering
   \includegraphics[width=0.80\textwidth]{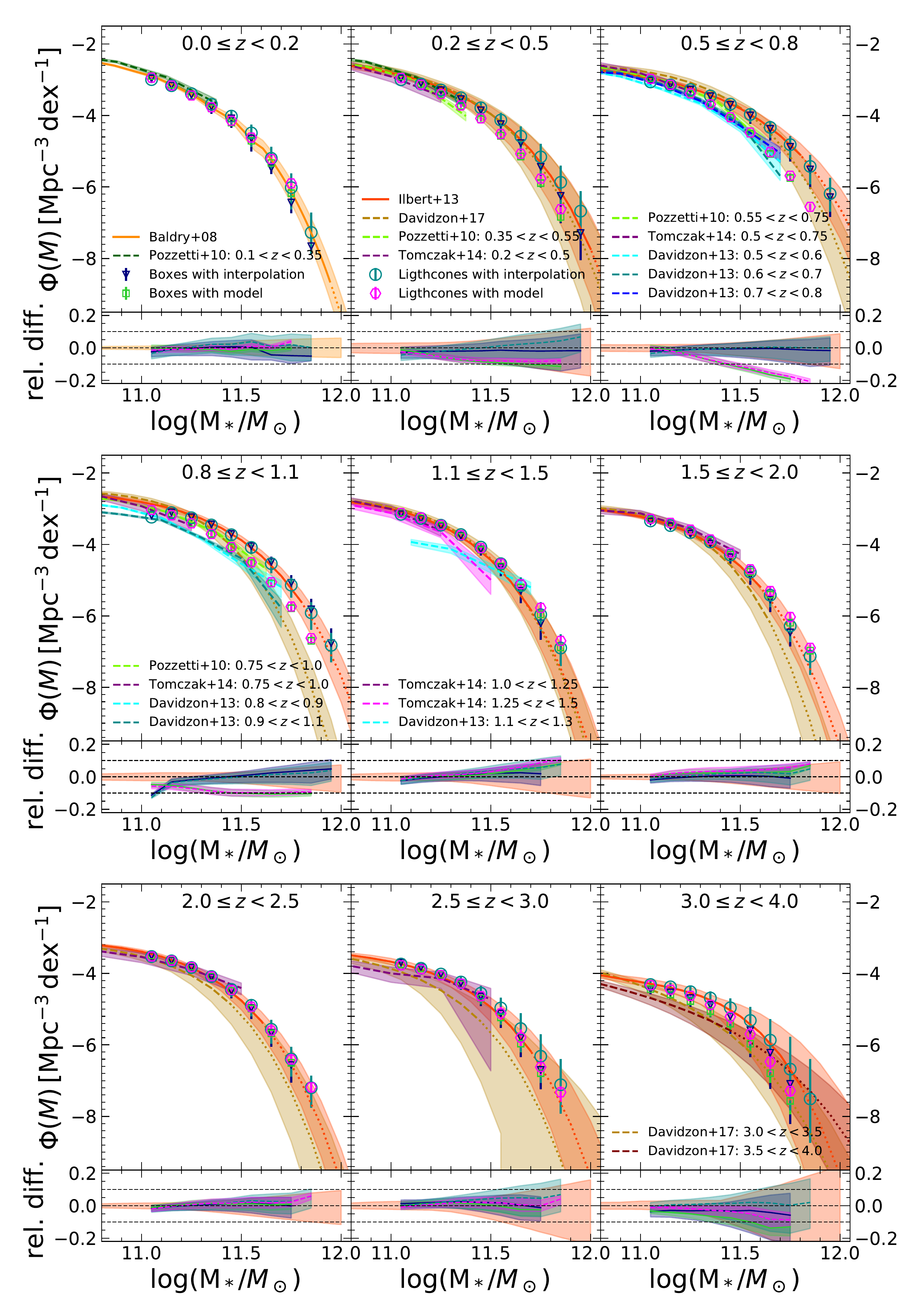}
  \caption{Differential SMFs in different redshift bins. Points represent the SMFs of the $\Lambda $CDM {\small DUSTGRAIN-}{\it pathfinder} simulation using the linear interpolation and  our reference model without scatter in comoving boxes and in light cones. Solid lines (and associated shaded regions) are the SMFs we used to calibrate our relation \citep{Ilbert13,Baldry08}. With dashed lines (and shaded regions) we show other literature results by \citet{Pozzetti10,Tomczak14,Davidzon13,Davidzon17}. As a general rule, the stellar mass at which the plotted best-fit Schechter functions turn from a continuos or dashed to a dotted line identifies the mass of the most massive observed galaxy. Moreover, when the redshift bin from other literature results is different from what is indicated at the top of each panel, the exact redshift binning is indicated in the inserted labels. Otherwise, if the redshift bins coincide, we report the references of the SMFs only once in the labels.
  The bottom panels show the relative difference between the results on the $\Lambda $CDM {\small DUSTGRAIN-}{\it pathfinder} simulation and the SMFs with which we calibrated the relation.}
\label{all_differentialSMF}
\end{figure*}

In this section, we investigate the accuracy of the SHMR we propose. To this aim, we assigned stellar masses to halos in comoving boxes and in light cones (using $M_{200}$ for main halos and $M_{\rm infall}$ for subhalos) of the $\Lambda $CDM {\small DUSTGRAIN-}{\it pathfinder} simulation and compared the results with observed SMFs.

\begin{figure}[ht]
   \centering
\includegraphics[width=0.49\textwidth]{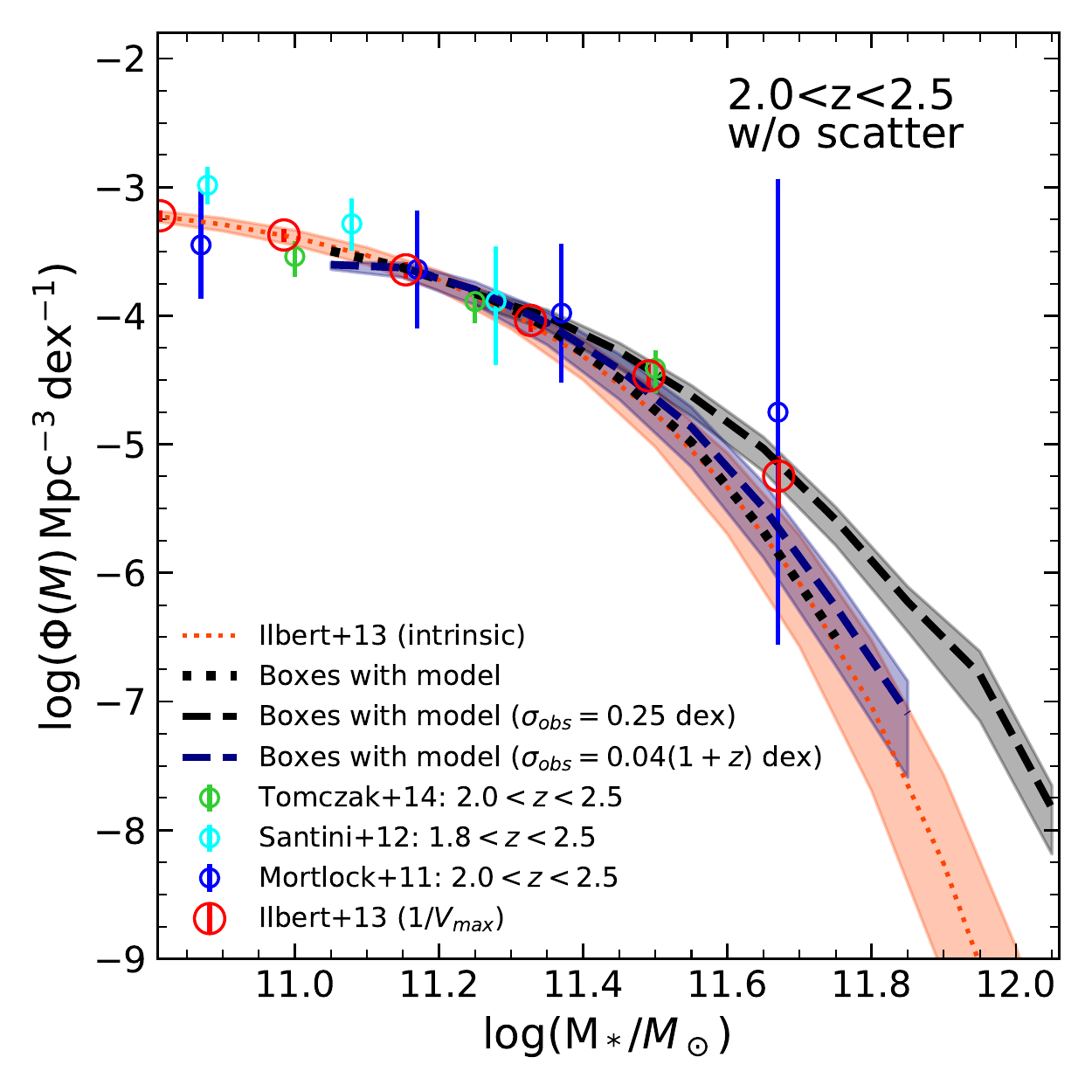}
  \caption{ Differential SMFs in the redshift range $2.0<z<2.5$ using our reference model without {\it \textup{relative}} scatter. Black dotted, black dashed, and blue dashed lines represent the SMF of the $\Lambda $CDM {\small DUSTGRAIN-}{\it pathfinder} simulation (calculated using the reference model on the comoving boxes), without {\it \textup{observational}} scatter, a fixed scatter (with $\sigma_{obs}=0.25$ dex), and a scatter that varies with redshift (with $\sigma_{obs}=0.04(1+z)$ dex), respectively. With open red circles we show the $1/V_{\rm max}$ points by \citet{Ilbert13} and with a dotted red line (and corresponding shaded area) their intrinsic best-fit Schechter function. With green, cyan, and magenta open points we show results from \citet{Tomczak14}, \citet{Santini12}, and \citet{Mortlock11}, respectively. }
              \label{Comparison_edd}
\end{figure}

\begin{figure}[ht]
   \centering
\includegraphics[width=0.49\textwidth]{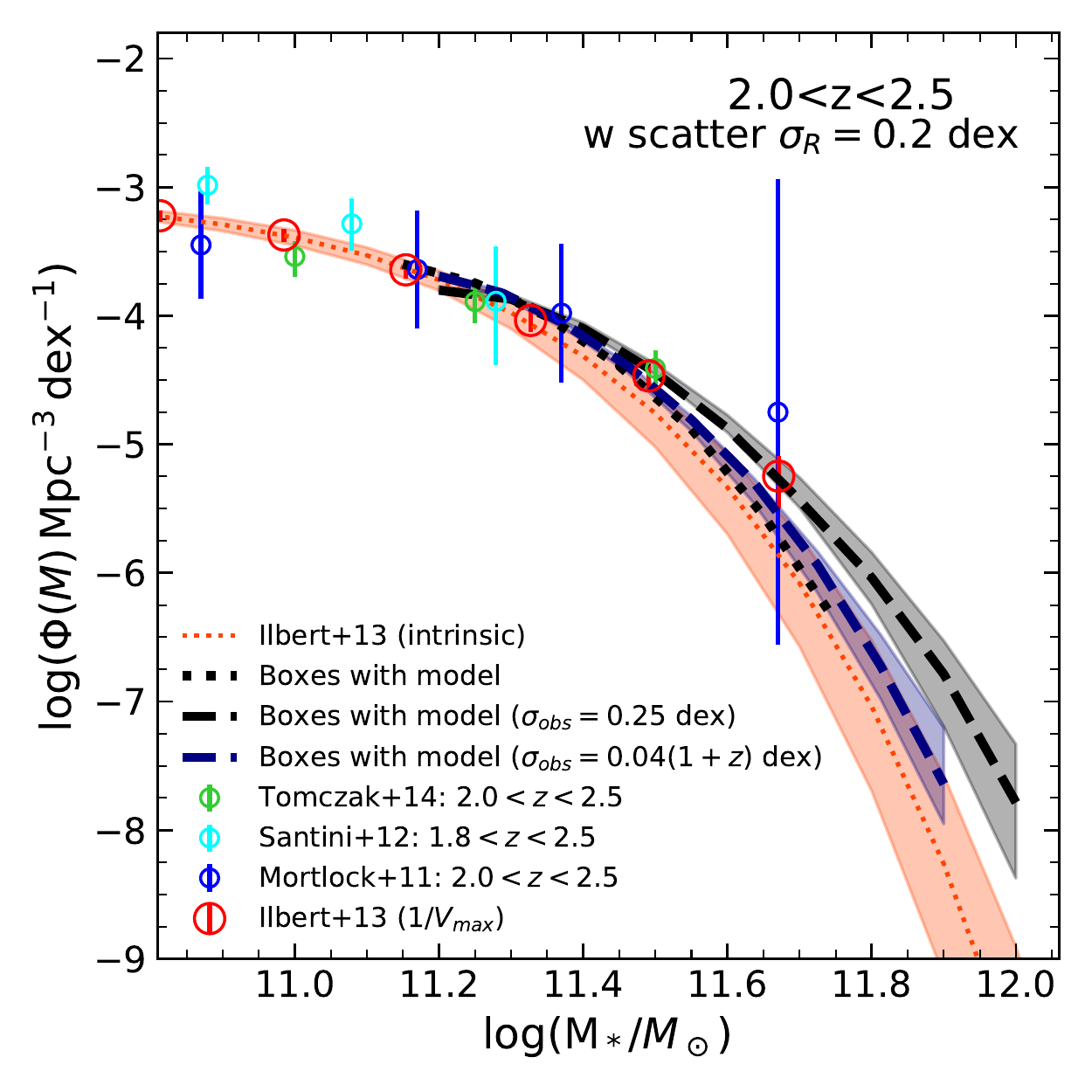}
  \caption{ Differential SMFs in the redshift range $2.0<z<2.5$ using the model with {\it \textup{relative}} scatter. Lines and symbols are the same as in Fig.~\ref{Comparison_edd}.}
              \label{Comparison_edd2}
\end{figure}

We used two methods to assign the stellar mass to simulated dark matter halos. The first consists of linearly interpolating between the best-fit results in individual redshift bins (see Table~\ref{tablefit} and Fig.~\ref{SHMR_fit}) between $M_h$, $M_*$ and redshift (we also used this to plot the colored layer in Fig.~\ref{SHMR3D}). In this way, we do not rely on any parameterization of the evolution for the parameters of the fit, but instead use the results of the direct comparison between the stellar and halo cumulative mass functions. This method is expected to (almost) perfectly reproduce, by construction, the \citet{Ilbert13} and \citet{Baldry08} SMFs. The second method consists of using the model proposed in Sect.~\ref{model}. This method has the advantage that it employs a simple functional form to assign the stellar mass to dark matter halos (and therefore is computationally very cheap). However, the parameters depend on redshift, as described by Eqs.~\ref{1}, \ref{2}, \ref{3}, and \ref{4}. For this reason, the model may not perfectly reproduce the observed SMFs we used to calibrate the SHMR at every redshift because these SMFs are evaluated on the SDSS and COSMOS fields, which have their specific features. 
We assigned stellar masses to halos in light cones and comoving boxes of the simulation using both methods. The described procedure was performed using results with and without { \it \textup{relative}} scatter. To generate the mock galaxy catalog in the latter case, we first assigned the stellar mass according to the corresponding SHMR (i.e., using the results detailed in Tables \ref{indfit_scat} and \ref{table:2}), and then we scattered the assigned stellar masses using a log-normal distribution with standard deviation $\sigma_R=0.2$ dex.
After the stellar mass was assigned to simulated dark matter halos, we calculated the SMFs only for stellar masses $M_* \geq  10^{11}\,M_\odot$  because of the mass resolution of the simulation, which is complete down to halo mass $M_h \geq M_{min,halo} = 10^{12.5}\,M_\odot$. The two cases we explored, with and without {\it \textup{relative}} scatter, agree well (see, e.g., Fig. \ref{Comparison_edd} and \ref{Comparison_edd2}, respectively). 
In both cases, it is evident that the models reproduce observations with good accuracy, with only small differences among them.
In Fig.~\ref{all_differentialSMF} we show the results for our {\it \textup{reference}} case without scatter in nine different redshift bins along with the SMFs on which the SHMR is calibrated (i.e. \citealt{Baldry08} and \citealt{Ilbert13}) and also other literature results by \citet{Pozzetti10}, \citet{Davidzon13}, \citet{Tomczak14}, and \citet{Davidzon17}. The \citet{Tomczak14} SMFs were estimated from the ZFOURGE survey \citep{Straatman16}, which includes three pointings in CDFS \citep{Giacconi02}, COSMOS, and UDS \citep{Lawrence07} fields. The total area used to evaluate SMFs by \citet{Tomczak14} is $\sim 316 \,\mathrm{arcmin}^2$ from $z=0.2$ to $z=3.0$. \citealt{Davidzon17} SMFs were evaluated in the COSMOS field following \citet{Ilbert13}, but with a different data release \citep{Laigle16}. Moreover, \citet{Davidzon17} restricted their analysis to a smaller area than \citet{Ilbert13} (i.e., the ultra-deep stripes covered by ULTRA-Vista with an area of $0.62\,\mathrm{deg}^2$) but with deeper observations ($K_s=24.7$ compared to $K_s=24$ of \citet{Ilbert13}), which in turn can increase the cosmic variance in \citet{Davidzon17}. This and the different SED-fitting procedure lead to the differences between the two works that are visible in Fig.~\ref{all_differentialSMF} (especially in the redshift bin $0.8\leq z<1.1$). The
\citet{Pozzetti10} SMFs were derived on the COSMOS field as well, but with the zCOSMOS spectroscopic survey data \citep{Lilly07}, which cover $\sim 1.4\,\mathrm{deg}^2$ up to $z\sim 1$. In this case, redshifts were estimated from spectroscopy, and stellar masses were derived by SED-fitting of the multiband photometry. Finally, the \citet{Davidzon13} SMFs were evaluated using the spectroscopic VIMOS Public Extragalactic Redshift Survey (VIPERS) \citep{Guzzo13,Scodeggio2018} covering $\sim 10\,\mathrm{deg}^2$ from $z\sim 0.5$ to $z\sim 1.1$.
When available (i.e., \citealt{Ilbert13,Davidzon17}) we used the best-fit intrinsic Schechter function, and in the other cases (i.e., \citealt{Pozzetti10,Davidzon13,Tomczak14}) we show the $1/V_{\rm max}$ estimates.

The bottom panels of Fig.~\ref{all_differentialSMF} show the relative differences between the results on the $\Lambda $CDM {\small DUSTGRAIN-}{\it pathfinder} simulation and the SMFs with which the SHMR was calibrated (i.e., \citealt{Baldry08} and \citealt{Ilbert13}).
The results for the interpolation and the model agree well with the observations with which we calibrated the relation, using either comoving boxes or light cones. In all cases (with the exception of the redshift bin $0.5 < z < 0.8$ using the model), the differences with the respect to the observed SMFs using the model (interpolation) are smaller than $\sim 10\%$ ($\sim 5\%$) of the value of the observed SMFs. The exception of redshift bin $0.5 < z < 0.8$, where the model differs by $\sim 15-20\%$ (at maximum) with respect to observations, is due to the cited overdensity in the COSMOS field. When we compared this with other observations (i.e., \citealt{Davidzon13,Davidzon17}), the differences in redshift bin $0.5 < z < 0.8$ are much less evident. In general, our results agree with the observed SMFs at all redshifts. 

We point out that our results depend on the SMFs we used to calibrate the relation, and if the calibrating SMFs change, the SHMR changes accordingly. In a future work we will perform the same analysis using different SMFs and extend it to even higher redshifts. Nonetheless, SDSS and COSMOS are currently the best fields with a homogeneous redshift coverage from $z=0$ to $z=4$ because of their statistical and photometric accuracy in the photometric redshifts, stellar masses, and SMF determination.

Up until now, we have derived the intrinsic SHMR because we adopted the intrinsic SMF, that is, deconvolved for the Eddington bias. Adding an additional scatter to mimic observational errors is instead essential to build realistic mock catalogs that take the errors ($\sigma_{obs}$) in the mass determination from photometry into account and can be directly compared with the observed SMF.
For each halo and each case (i.e., with and without { \it \textup{relative}} scatter), we now assigned a stellar mass following our intrinsic SHMRs, and convolved these masses with a log-normal distribution with standard deviation $\sigma_{obs}$. We explored the effects of using two different values for the standard deviation. 
We assumed the first value to be independent of the halo mass but dependent on redshift. Following COSMOS results \citep{Ilbert13,Davidzon17}, we assumed $\sigma_{obs}=0.04(1+z)$\,dex, that is, the scatter for a COSMOS-like dataset with which we calibrated our method. 
The second value was assumed to be constant with halo mass and redshift, $\sigma_{obs}=0.25$\,dex, following the results by  \citet{Mortlock11} at $1<z<3$ on the GOODS survey.

We then calculated the $\phi_{\rm convolved}$ using the two different scatters and compared this to literature results.  In Figs.~\ref{Comparison_edd} and \ref{Comparison_edd2} we show as examples the convolved SMFs we derived using the models (with and without {\it \textup{relative}} scatter) from the $\Lambda $CDM {\small DUSTGRAIN-}{\it pathfinder} simulation boxes in one redshift bin, but the same effects also apply to the other redshift bins.
We also show the observed $1/V_{\rm max}$ SMF on the COSMOS field from \citet{Ilbert13} and its corresponding intrinsic SMF. In addition, we show other observed SMFs by \citet{Mortlock11}, \citet{Santini12}, and \citet{Tomczak14}. The first was derived in the Great Observatories Origins Deep Survey (GOODS) with the Near Infrared Camera and Multi-Object Spectrometer (NICMOS) \citep{Conselice11} on board the Hubble Space Telescope ($H_{160}=26.8$ over $\sim 43.7$\,arcmin$^{2}$). The second, \citet{Santini12}, is a study in the GOODS-S field on a relatively small but deep field ($K_s=25.5$, over $\sim 33$\,arcmin$^{2}$). 

The effect of observational errors on the stellar masses with a COSMOS-like scatter (i.e., $\sigma_{obs}=0.04(1+z)$) is quite small and does not affect the derived SMF strongly for either case, whose values are both in agreement with the \citet{Ilbert13} results on the COSMOS field. 
In other surveys, the photometric errors, which affect photometric redshift and mass determination, might be larger and should be taken into account when a realistic mock catalog is built. For example, we compared our SMF forecasts, obtained with a fixed scatter value, with other literature results, such as \citet{Mortlock11} and \citet{Santini12}. 
 In the case of a fixed scatter, $\sigma_{obs}=0.25$, our convolved SMF also reproduces the excess fairly well with respect to the intrinsic SMF that is observed at high masses in the SMFs as a result of the Eddington bias. 

\section{Clustering}\label{clustering}
After populating halos with galaxies and verifying that the SMF was correctly reproduced, we measured the galaxy angle-averaged two-point correlation function (2PCF), $\xi(r)$, to evaluate whether the SHMR we propose is able to reproduce the observed clustering of galaxies as well. We performed this calculation on two mock catalogs that were constructed from the boxes of the $\Lambda $CDM {\small DUSTGRAIN-}{\it pathfinder} simulation. In the first we used our reference model without scatter, and in the second we considered the model with {\it r\textup{elative}} scatter. Because we wished to compare our results with observations, an observational scatter was also added (to both mocks) with a log-normal distribution with standard deviation $\sigma_{obs}=0.04(1+z)$\,dex. All the measurements were made with the \textit{CosmoBolognaLib} \citep{Marulli16}, a set of {\it \textup{free software}} C$++$/Python numerical libraries for cosmological calculations. The 2PCF was measured on the comoving boxes for galaxies with $\log(M_*/M_\odot)\geq 11$, using the \citet{Landy93} estimator, which has been shown to provide a nearly unbiased estimate of the 2PCF, while minimizing its variance \citep[see, e.g.,][]{Keihanen2019},
\begin{equation}
\hat{\xi}(r) \, = \, \frac{N_{RR}}{N_{GG}} \frac{GG(r)}{RR(r)} -2
  \frac{N_{RR}}{N_{GR}} \frac{GR(r)}{RR(r)} +1 \,,
  \label{eq:xiLS}
\end{equation}
where $GG(r)$, $RR(r),$ and $GR(r)$ are the binned numbers of galaxy-galaxy, random-random, and galaxy-random pairs with distance $r \pm\Delta r$, while $N_{GG}=N_C(N_G-1)/2$, $N_{RR}=N_R(N_R-1)/2$, and $N_{GR}=N_GN_R$ are the total numbers of galaxy--galaxy, random--random, and galaxy--random pairs, respectively. The random samples were constructed to be five times larger than the galaxy samples.

\begin{figure}[ht]
   \centering
   \includegraphics[width=9cm]{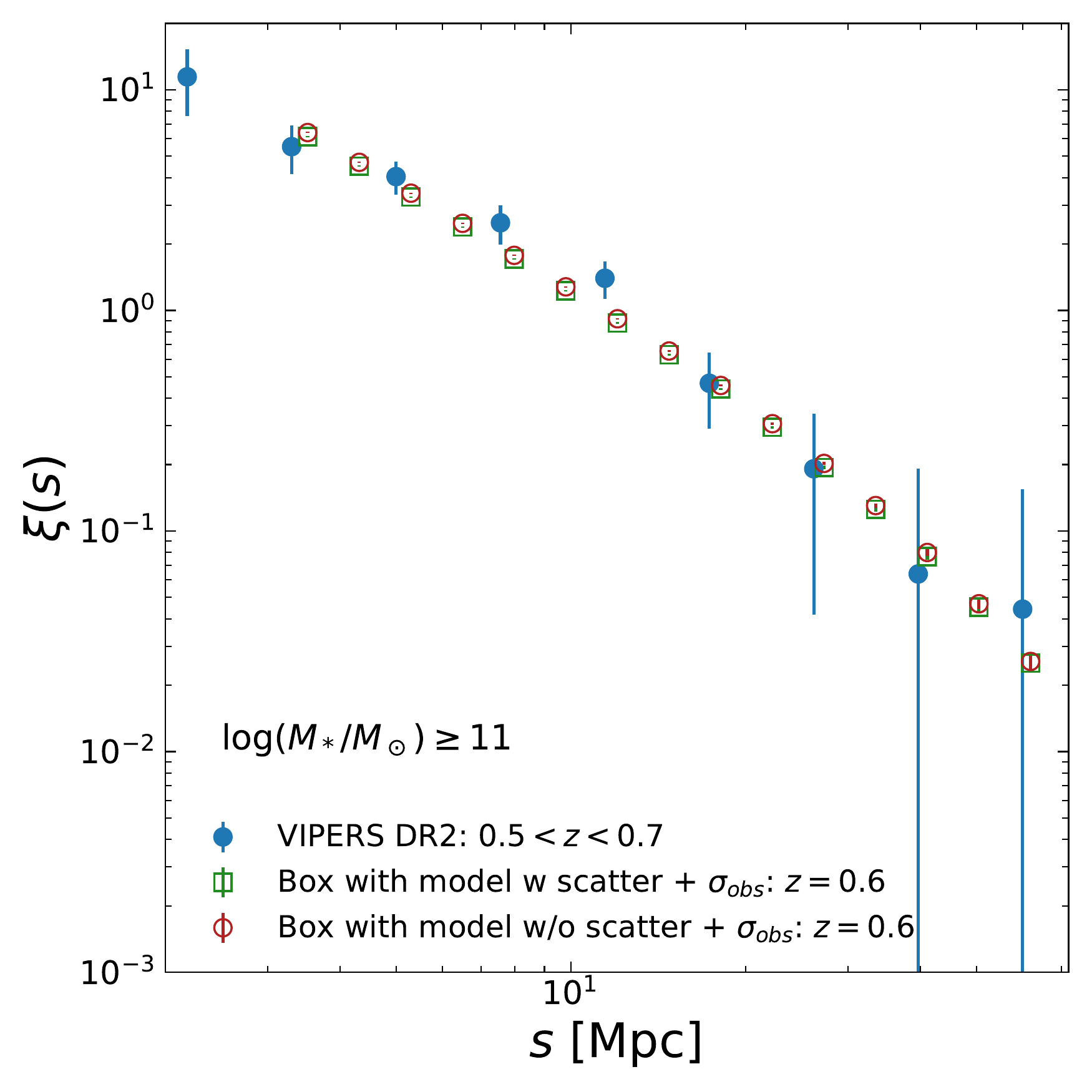}
   \includegraphics[width=9cm]{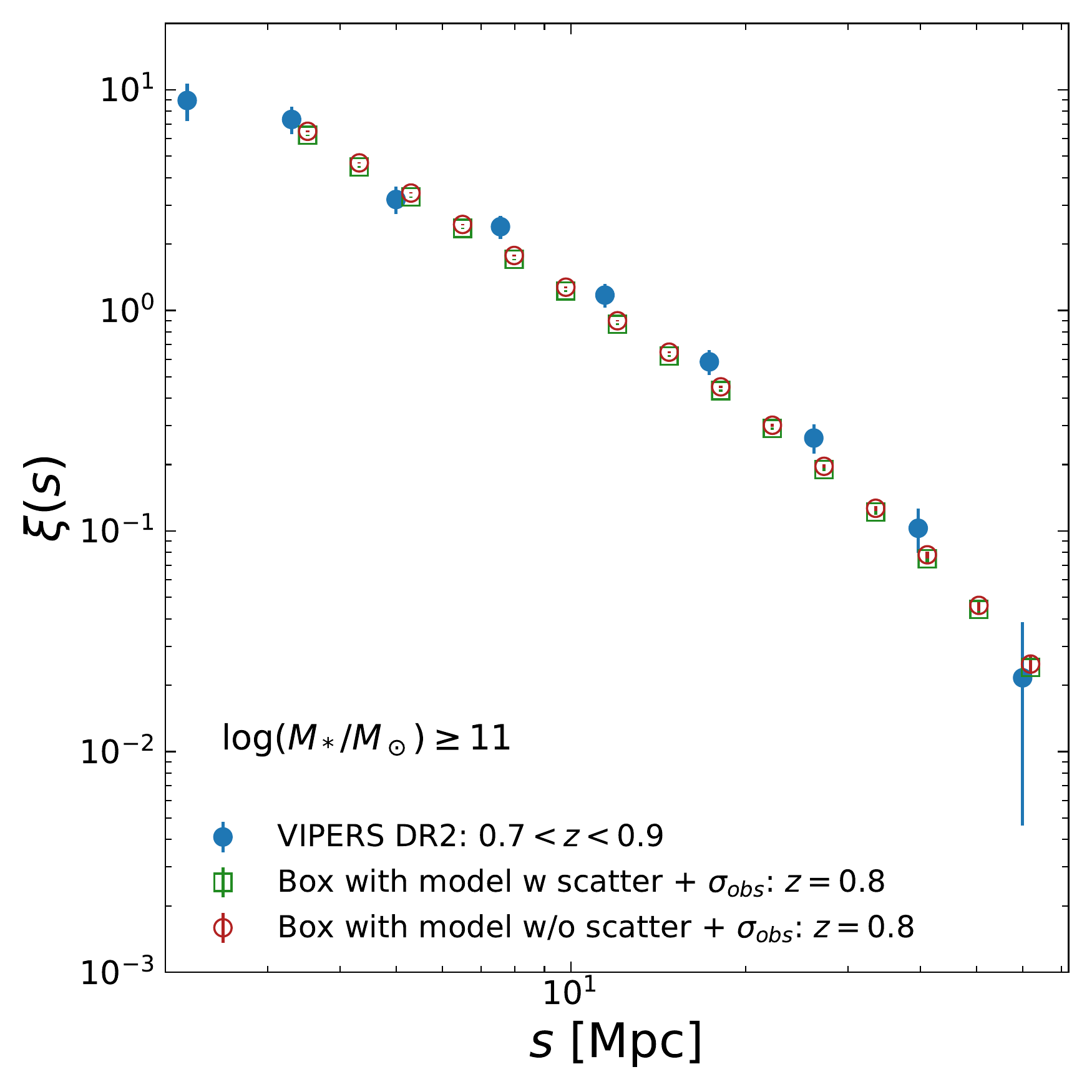}
  \caption{2PCF of galaxies with $M_* \geq  10^{10.11}\,M_\odot$. Open red circles and green squares represent the 2PCF of mock galaxies with stellar masses evaluated using the model (with and without scatter) at $z=0.6$ (top) and $z=0.8$ (bottom). Filled blue points show the 2PCF of the VIPERS galaxies in the redshift bins $0.5<z<0.7$ (top) and $0.7<z<0.9$ (bottom).}
  \label{clust1}%
\end{figure}

 We computed the 2PCF from a minimum scale of $s_{\rm min}=0.22\,$ Mpc $(=0.15\,h^{-1}$\,Mpc) to a maximum scale of $s_{\rm max}=104\,$Mpc $(=70\,h^{-1}$\,Mpc) over $30$ bins that were equally separated in logarithmic scale. The 2PCF uncertainties were estimated with the bootstrap method by dividing the datasets into $27$ subsamples, which were then resampled in $100$ datasets with replacements, measuring the 2PCF in each of them \citep{Norberg2009}.
We translated the real-space galaxy 2PCF ($\xi(s)$) into the redshift-space galaxy 2PCF ($\xi(r)$) as follows \citep{Kaiser1987}:
\begin{equation}
\xi(s) = \left(1+ \frac{2\beta}{3} + \frac{\beta^2}{5}\right)\xi(r) \; ,
\label{eq:Kaiser}
\end{equation}
where $\beta\equiv f(z)/b(z)\simeq\Omega_{\rm M}^{0.545}(z)/b(z)$ is the linear growth rate \citep[see, e.g.,][]{Marulli17}. The linear bias, $b(z)$, was estimated from the square root of the ratio between the real-space 2PCFs of galaxies and dark matter at the largest probed scales (i.e., $s > 15\,\mathrm{Mpc}=10\,h^{-1}\,\mathrm{Mpc}$). To compute the latter, we used a Fourier transform on the nonlinear matter power spectrum computed with {\small CAMB}, including {\small HALOFIT} \citep{Lewis00,Smith03}. 

 The Kaiser model in Eq.~\eqref{eq:Kaiser} does not provide
  an accurate description of the redshift-space 2PCF at small scales,
  that is, in the so-called one-halo regime. However, as we described above, we
  are complete in subhalos only down to $M_h=10^{12.5}M_\odot$, thus
  our clustering measurements are only reliable at large scales
  ($\gtrsim 3$ Mpc), where the Kaiser model is reliable. To determine whether the lack of the one-halo term in the 2PCF of our mock galaxy
  catalogs was caused by the adopted SHAM prescriptions, we repeated
  the analysis by measuring the 2PCF of the full-halo catalogs, without
  mass selection.  We found no sign of the one-halo
  term in this case either. It can therefore be ascribed to the resolution of the
  simulation.

  An alternative way to compare the spatial properties of our mock
  galaxy catalogs to real data is through the projected
  clustering. However, we preferred not to calculate the projected 2PCF
  in this analysis and retained the angle-averaged 2PCF in
  redshift space to avoid additional uncertainties that might be
  caused by the numerical methods for constructing redshift-space mock
  samples from the available discrete set of simulation
  snapshots. Moreover, the 2PCF would have to be integrated along the
  line of sight from the smallest scales, where our predicted 2PCF is
  not reliable because we are incomplete in low-mass subhaloes. 

Figure \ref{clust1} shows our results at $z = 0.6$ and $0.8$ for scales $s\geq 3$ Mpc compared to the observed galaxy 2PCF from the final release of the VIPERS survey \citep{Guzzo13, Scodeggio2018} in the redshift bins $0.5 < z < 0.7$ and $0.7 < z < 0.9$. The analysis was restricted to galaxies with $M_* \geq  10^{11}\,M_\odot$ to be complete in the mock catalogs.
Fig. \ref{clust1} shows that the normalization of the 2PCF is slightly lower when the scatter is included in the model. This is expected because in this case, galaxies are hosted on average by lower mass halos because of the shape of the MF. These lower mass halos are characterized by a lower bias.

The 2PCF of the VIPERS galaxies is measured similarly to what has been reported by \citet{Marulli13}. 
The random sample we used to estimate the 2PCF of the VIPERS galaxies was constructed as in \citet{Marulli13}, with $30$ times more objects than galaxies. The weighting scheme we adopted has been described by \citet{Pezzotta17} and \citet{delaTorre2017}. 
The errors on the VIPERS 2PCF were estimated with the bootstrap method in the same way as for the mock catalogs. 

Overall, the 2PCFs of our mock galaxies appear to agree well with the VIPERS galaxies at all redshifts for scales $s\gtrsim 3$ Mpc, that is, in the two-halo regime, where the 2PCF is dominated by pairs of galaxies that are hosted in two different dark matter halos. To assess the statistical level of the comparison, the analysis needs to properly consider the full covariance matrices of the measurements. This is beyond the scope of this work. 

At smaller scales our results disagree with VIPERS measurements. As described above, this is due to the resolution of the considered simulation. The number of subhalos with $\log(M_h/M_\odot)<12.5$ that are expected to host massive galaxies is not resolved with a sufficient number of dark matter particles. This biases the number of pair galaxies that are hosted in the same halo that contribute to the one-halo clustering term at small scales.  

\section{Summary and conclusions}

The main results and conclusions of the present study can be summarized as follows :
\begin{enumerate}
\item We presented a new SHMR based on a (sub-) halo abundance matching technique assuming a deterministic relation between halo and stellar mass. We adopted the \citet{Despali16} halo mass function, after showing that it is fully consistent (at all redshifts and halo masses) with our reference dark matter simulation ($\Lambda $CDM {\small DUSTGRAIN-}{\it pathfinder} simulation, \citealt{Giocoli18}) using subhalos with infall mass or observed mass, and using comoving boxes or light cones. We combined the HMF with \citet{Baldry08} and \citet{Ilbert13} {\it \textup{intrinsic}} SMFs to calculate the SHMR from $M_*\sim10^8\,M_\odot$ (i.e., $M_h\sim10^{10.5}\,M_\odot$) to $M_*\sim10^{12}\,M_\odot$ (i.e. $M_h\sim10^{15}\,M_\odot$), homogenously from $z=0$ to $z=4$, in nine different redshift bins. We also presented results when a {\it \textup{relative}} scatter in stellar mass at fixed halo mass was introduced in the relation to account for different halo accretion histories, spin parameters, and concentrations. This was made by adding a scatter drawn from a log-normal distribution with a standard deviation $\sigma_R=0.2$ dex to the model SMF when we compared it to the {\it \textup{intrinsic}}  SMF. We showed that this scatter mostly affects the high-mass end of the SHMR. In addition, we find that results with and without relative{\it } scatter are consistent between each other.
\item We proposed a simple model that parameterizes the evolution in redshift of the SHMR with and without {\it \textup{relative}} scatter, following the formalism presented in \citet{Moster10}. The model tends to smooth all peculiar features that characterize the SMFs we used to calibrate the SHMR. The model therefore predicts a smooth evolution with redshift.
\item When a constant baryon fraction is assumed, the SHMR can be interpreted as the SFE. We find that it monotonically increases as a function of halo mass (with a slope $\beta\sim 1$ almost constant with redshift) with a peak at $M_h\sim 10^{12}M_\odot$, and a change  in slope ($\gamma \sim 0.7 - 0.6$) that decreases monotonically at higher halo masses. Moreover, we note that above the characteristic halo mass peak, $M_A=M_{h,{\rm peak}}$, the SHMR increases with increasing redshift. At masses below the characteristic halo mass, this trend is reversed. This means that massive galaxies (i.e., halo masses higher than the SHMR peak) form with a higher efficiency at higher redshifts. In contrast, low-mass galaxies (corresponding to halo masses lower than the SHMR peak) formed with higher efficiency at lower redshifts. These trends, which are also preserved when the baryon fraction depends on halo mass or redshift, are a manifestation of the downsizing effect.

\item We compared our results with several methods to link dark matter halos to galaxies (e.g., HOD, CLF, and EM) that are available in the literature. While at $z\sim 0$ all works agree fairly well (within a factor of $\sim 3$), especially in the slopes of the SHMR, at $z>0,$ the results of different studies show large differences not only in the normalization, but also in the slopes of the relation. These differences can be ascribed to the different SMFs that were used to calibrate the models and also to different halo-finder algorithms that were used to derive the halo masses.
\item Our results were also compared to several SAMs. In this case, the differences are quite large (up to a factor of $\sim 8$) in the normalization and in the slopes, which reflects the difficulty of treating the physical processes related to galaxy formation and evolution. In particular, we find that for increasing redshift, SAMs predict a less pronounced decreasing evolution of the SHMR at low halo masses and a decreasing trend at high halo masses as well. This is at odds with our observed increasing values. 
\item We studied the redshift evolution of the peak halo mass ($M_{h,{\rm peak}}$) and the corresponding $(M_*/M_{h})_{\rm peak}$ value. We find that $M_{h,{\rm peak}}$ increases with redshift, while $(M_*/M_h)_{\rm peak}$ remains approximately constant. We interpret the increasing value of $M_{h,{\rm peak}}$ with redshift as a signature of the downsizing effect, where $M_{h,{\rm peak}}$ is the halo mass corresponding to the most efficient star formation. Interpreting  $(M_*/M_h)_{\rm peak}$ as the peak of SFE, we find that it increases from $\sim 0.3$ at $z\sim 4$ to $\sim0.35$ at $z\sim 0$. These values could increase by a factor up to $\sim 2$ if we were to also consider the fraction of gas that is returned to ISM during stellar evolution.
\item We investigated the accuracy of the SHMRs we propose by assigning stellar masses to halos in the $\Lambda $CDM {\small DUSTGRAIN-}{\it pathfinder} simulation (complete for $M_h>10^{12.5}M_\odot$) and compared the results with literature SMFs. Our results using the interpolation between the best-fit results in individual redshift bins agree excellently well with observations; the differences are always smaller than $\sim 5\%$ when boxes or light cones are used at all redshifts. With the model, differences are always smaller than $\sim 10\%$ at all redshifts. In the redshift bin $0.5 < z < 0.8$ the model differs by $\sim 15-20\%$ (at maximum) with respect to observations because of an overdensity in the COSMOS field. When we compare our results with different observations, with different datasets, and on different fields, the differences are much less evident.
\item We also studied the clustering as a function of stellar mass. We evaluated the 2PCF in the redshift space and compared the results with observed results from the VIPERS survey. We find a good agreement from $z\sim 0.5$ to $z\sim 1.1$ for massive galaxies ($M_*>10^{11}M_\odot$) for the two-halo regime (i.e., for scales $s\gtrsim 3$ Mpc).
\end{enumerate}

To conclude, from our analysis we found that the process of star formation (i.e., the process of turning gas into stars) is remarkably inefficient (with peak efficiencies $<35\%$ at all masses and redshifts). We showed that current SAMs do not reproduce our results well, which indicates that several physical mechanisms of galaxy formation and evolution are still not well understood.  
On the observational side more work is also needed in order to understand the scatter between different literature results, especially at $z>0$.

Studies of the galaxy-halo connection are essential to interpret galaxy surveys and to provide key insights into the problems of galaxy formation and evolution. The model we propose, applied to N-body dark matter simulations, is able to reproduce fairly good observed SMFs and clustering measurements of massive galaxies down to the resolution limit of the simulation. The good overall agreement indicates a realistic assignment of galaxy masses into halos. This can be used to create realistic galaxy mocks for future surveys or missions and to populate the N-body dark matter simulations with galaxies. 

The next generation of surveys, such as Euclid, are likely to shed light on the galaxy–halo connection over mass, redshift, and environment. This will provide major improvements in our understanding of galaxy formation, cosmological parameters, and the nature of dark energy and neutrino mass.
At the same time, the precision of models for the galaxy–halo connection need to be improved in order to keep up with the pace of the data.

\begin{acknowledgements}
We thank Iary Davidzon, Paola Santini, Peter Behroozi and Alexander Knebe for providing their data in electronic form. Data compilations for other studies used in this paper were made much more accurate and efficient by the online WEBPLOTDIGITALIZER code. We thank the referee for the useful comments. GG, LP, MB, CG, FM acknowledges the grants
ASI n.I/023/12/0, ASI-INAF n. 2018-23-HH.0 and PRIN
MIUR 2015 “Cosmology and Fundamental Physics: illuminating the Dark Universe with Euclid”. 
\end{acknowledgements}

\bibliographystyle{aa} 
\bibliography{SHMR}
\begin{appendix}

\section{Details on the literature}\label{appendix1}

In this appendix we describe the literature works with which we compare our results in Figs.~\ref{SHMR_HOD_z0} and \ref{SHMR_HOD_z13}.
The works that use a very similar method to ours are \citet{Moster10,Moster13} and \citet{Behroozi10,Behroozi13}, even though all cited works show some differences in the results with respect to ours at all redshifts.
As may be expected, it is much harder to directly measure the SHMR at $z>0$. This results in relatively fewer published results to which we may our results compare.

\subsection{\cite{Moster10,Moster13}}

\citet{Moster10} and \citet{Moster13} used the same functional form as we adopted for the SHMR. Both \citet{Moster10} and \citet{Moster13} used a SHAM technique to match simulated halos from the Millennium Simulation \citep{Springel05} to observed SMFs. However, they used different SMFs based on the results of \citet{Panter07} at $z=0$, and \citet{Fontana06} at $z>0$ in \citet{Moster10}, and \citet{Li&white09} at $z\sim 0$ and \citet{Perez08} at $z>0$ in \citet{Moster13}. This choice is reflected in differences in the stellar mass at fixed halo mass. 

At $z\sim 0$ our results agree well with \citet{Moster13} at high masses (i.e., $\log(M_h/M_\odot) \gtrsim 13.0$) and low masses (i.e., $\log(M_h/M_\odot) \lesssim 11$), while small differences are present at intermediate halo masses (e.g., a difference of a factor of $\sim 1.2$ in $(M_*/M_h)$ is found at $\log(M_h/M_\odot)=12.3$). At increasing redshift the agreement worsens. At $z\sim1$ and $z\sim 3$ our results differ not only in the SHMR normalization, but also in the slopes at high and low masses (at $\log(M_h/M_\odot)=12.3$ a difference of a factor of $\sim1.5$ and $\sim 1.4$ in $(M_*/M_h)$ is seen at $z\sim1$ and $z\sim3,$ respectively). The \citet{Moster10} results differ from ours at any redshift in the values of the stellar mass at fixed halo masses (with a difference of a factor of $\sim1.2$ in $(M_*/M_h)$ at $\log(M_h/M_\odot)=11.8$ at $z=0$, which increases to a factor of $\sim 1.7$ at $\log(M_h/M_\odot)=12.5$ at $z\sim 3$), but also in the low- and high-mass slopes of the SHMR.

\subsection{\cite{Guo10}}

\citet{Guo10} used an approach similar to that of \citet{Moster10}, based on stellar masses from \citet{Li&white09}, but they did not account for scatter in stellar mass at fixed halo mass. Consequently, their results do not match \citet{Moster10}.
As for \citet{Moster13}, at $z\sim 0$, our results agree well at high masses (i.e., $\log(M_*/M_h) \gtrsim 13.0$) but show differences at lower halo masses (e.g., a difference of a factor of $\sim1.2$ in $(M_*/M_h)$ is found at $\log(M_h/M_\odot)=11.8$).

\subsection{\cite{Zheng07}}

\citet{Zheng07} used the galaxy clustering for luminosity-selected samples in the SDSS (at $z\sim 0$) and in the DEEP2 Galaxy Redshift Survey (\citealt{Coil06} at $z\sim 1$) to constrain the halo occupation distribution (HOD). This gives a direct constraint on the $r$-band luminosity of central galaxies as a function of halo mass. Stellar masses for this sample were then determined using the $g - r$ color and the $r$-band luminosity, assuming a WMAP1 cosmology. This method allows for scatter in the luminosity at fixed halo mass to be constrained as a parameter in the model. Results from \citet{Zheng07} are in fairly good agreement with our results at $z\sim 0$, while the difference is larger (up to $0.5$\,dex) at $z\sim 1$.  

\subsection{\cite{Behroozi10,Behroozi13}}

\citet{Behroozi10,Behroozi13} used an approach similar to what we used here, but with a much more complicated functional form to parameterize the SHMR (also accounting for a variable scatter in the relation for a total of $14$ free parameters in the fit, compared to the $4$ we used). Moreover, at $z\sim 0$ \citet{Behroozi10} used \citet{Li&white09} SMFs, and \citet{Behroozi13} used \citet{Moustakas13}. At higher redshifts,  \citet{Behroozi10} and \citet{Behroozi13} used several SMFs (see their papers for details).
For both works, at $z\sim 0$ the slope of the SHMR is similar to ours, but the normalization in $(M_*/M_h)$ is different up to a factor of $\sim1.4$ and $\sim1.6$ at $\log(M_h/M_\odot)=11.8$ for \citet{Behroozi10} and \citet{Behroozi13}, respectively. At $z\sim 1$ and $z\sim 3$ there is not agreement on the slopes of the SHMR between Behroozi's results and ours. At $z\sim 1$, differences on the normalization increase to a factor of $\sim2$ and $\sim 1.4$ at $\log(M_h/M_\odot)=11.8$, while for $z\sim 3$ we find differences of a factor of $\sim 1.2$ and $\sim 1.4$ at $\log(M_h/M_\odot)=12.5$ for \citet{Behroozi10} and \citet{Behroozi13}, respectively.

\subsection{\cite{Reddick13}}

\citet{Reddick13} used additional input from the correlation function and conditional SMF as measured by SDSS. This result is fairly different from our result in the normalization of the SHMR. In particular, a difference of a factor of $\sim1.8$ in $(M_*/M_h)$ is found at $\log(M_h/M_\odot)=11.8$.

\subsection{\cite{Yang12}}

\citet{Yang12} adopted an approach similar to the abundance matching, but with some differences. In particular, their approach assumes the subhalo abundance as a function of the mass at accretion and the accretion time, following the \citet{Yang11} model. At $z\sim 0$ they used SMFs of SDSS DR7 \citep{Abazajian09}, while for $z>0$ they used SMFs from \citet{Perez08} and \citet{Drory05}. At $z\sim 0$, their results are fully consistent with ours down to halo masses of $\log(M_h/M_\odot)\sim12$. At lower masses, \citet{Yang12} predict higher stellar masses with respect to our result. At $z>0,$ the agreement becomes worse and the  differences increase to a factor of $\sim 2.7$ at $\log(M_h/M_\odot)=12.5$ in $(M_*/M_h)$ at $z\sim 3$.

\subsection{\cite{Rodriguez17}}

\citet{Rodriguez17} estimated the SHMR as a function of redshift using a SHAM technique and several observed SMFs, based on different observational campaigns and techniques (see \citealt{Rodriguez17} for details). Their results are not consistent with the results we obtain, with differences in the normalization of the SHMR and also in the slope of the relation at low and high masses. A difference of a factor of $\sim1.4$ in $(M_*/M_h)$ is found at $\log(M_h/M_\odot)=11.8$ between their results and ours.

\subsection{\cite{Carretero15}}

\citet{Carretero15} combined a HOD model to a SHAM technique to link galaxy luminosities to dark matter halos. In particular, they used dark matter simulations from MICE \citep{Fosalba08} along with luminosity functions from the SDSS survey \citep{Blanton03}. Even if the slope of the SHMR at high halo masses (i.e., $\log(M_h/M_\odot)\gtrsim 12.5$) is similar to the slope in our work, we find a large discrepancy in the normalization of the SHMR, with a difference in $(M_*/M_h)$ of a factor of $\sim 2$ at $\log(M_*/M_h)=11.8$. 

\subsection{\cite{Behroozi19}}

\citet{Behroozi19} presented a method for determining individual galaxy star formation rates from their host halos potential well depths, assembly histories, and redshifts. For each halo, the galaxy stellar mass is derived from the star formation histories along the halo assembly and merger history. The model is calibrated through several observations of the SMF and the cosmic star formation rate. However, their results disagree with ours at all redshifts, with differences in the slopes (at high and low halo masses) and in the normalization of the SHMR. At a halo mass of $\log(M_h/M_\odot)=11.8$ ($12.0$, $12.4$), a difference of a factor $\sim1.6$ ($\sim1.5$, $\sim1.6$) in $(M_*/M_h)$ is found at $z\sim0$ ($z\sim1$, $z\sim 3$).

\subsection{\cite{Moster18}}

 \citet{Moster18} also presented an empirical model of galaxy formation. They assigned a star formation rate to each dark matter halo based on its growth rate and computed the stellar masses by integrating it. Several observations of the cosmic star formation rate densities were used to calibrate the model (see \citet{Moster18} for more details). At $z\sim 0$, the \citet{Moster18} results are inconsistent with our findings in the slopes of the SHMR and in the normalization (e.g., a difference of a factor of $\sim1.5$ in $(M_*/M_h)$ can be seen at $\log(M_h/M_\odot)=11.8$). Moving to $z\sim 1$, we instead find good agreement in the slopes of the SHMR and also in the normalization at masses $\log(M_h/M_\odot)\lesssim 12.0$. At higher halo masses, the normalization differs (by a factor of $\sim1.2$ in $(M_*/M_h)$ at $\log(M_h/M_\odot)=12.3$). At $z\sim 3,$ our results disagree with those of \citet{Moster18}. Not only the slopes and normalization differ, but the halo mass at the peak of the SHMR is also remarkably different (see Sect.~\ref{peak} for a more detailed analysis of the SHMR peaks).

\subsection{\cite{Yang09}}

At the high-mass end, \citet{Yang09} have directly identified clusters and groups corresponding to dark matter halos, and measured the stellar masses of their central galaxies. They used a group catalog matched to halos to determine halo masses. Their results agree well with ours for halos at all masses.

\subsection{\cite{Wang&Jing10}}
\citet{Wang&Jing10} applied the empirical method built for redshift $z = 0$ in their previous work \citep{Wang06} to a higher redshift to link galaxy stellar mass directly with its hosting dark matter halo mass at redshift of about $0.8$. The SHMR is found by fitting the SMF and the correlation functions at different stellar mass intervals from VIMOS-VLT Deep Survey (VVDS) observation. Positions of galaxies are predicted by following the merging histories of halos and the trajectories of subhalos in the Millennium Simulation \citep{Springel05}.

\subsection{\cite{Legrand19}}
\citet{Legrand19} used a parametric abundance matching technique to link observed SMFs on the COSMOS field \citep{Davidzon17} to halo mass functions. In particular, they used the method developed in \citet{Behroozi10,Behroozi13}. Their results agree with ours at low redshift ($z\sim 0$) and high halo masses ($\log(M_h/M_\odot)\gtrsim 12.5$), but show differences up to $\sim 0.5$ dex at other halo masses and redshifts.
\end{appendix}

\end{document}